\documentclass[secnumarabic, graphics,floatfix, nofootinbib,tightenlines,nobibnotes, aps, prl, 12pt]{revtex4-1}
\usepackage{graphicx}
\usepackage[english]{babel}
\usepackage[utf8]{inputenc}
\usepackage{amsmath,amssymb}
\usepackage{amsfonts}
\usepackage{multirow}
\usepackage{pstricks}
\usepackage{float}
\usepackage{subfigure}
\usepackage{color}
\usepackage{epsfig}
\usepackage{pst-tools}

\begin{document}

\title{Hydrodynamic results on multiplicity fluctuations in heavy-ion collisions}

\author{Hong-Hao Ma$^{1,2}$}
\author{Dan Wen$^{1,2}$}
\author{Kai Lin$^{3,4}$}
\author{Wei-Liang Qian$^{4,2,1}$}
\author{Bin Wang$^{1}$}
\author{Yogiro Hama$^{5}$}
\author{Takeshi Kodama$^{6,7}$}

\affiliation{$^{1}$ Center for Gravitation and Cosmology, College of Physical Science and Technology,Yangzhou University, Yangzhou 225009, China}
\affiliation{$^{2}$ Faculdade de Engenharia de Guaratinguet\'a, Universidade Estadual Paulista, 12516-410, Guaratinguet\'a, SP, Brazil}
\affiliation{$^{3}$ Hubei Subsurface Multi-scale Imaging Key Laboratory, Institute of Geophysics and Geomatics, China University of Geosciences, 430074, Wuhan, Hubei, China}
\affiliation{$^{4}$ Escola de Engenharia de Lorena, Universidade de S\~ao Paulo, 12602-810, Lorena, SP, Brazil}
\affiliation{$^{5}$ Instituto de F\'isica, Universidade de S\~ao Paulo, C.P. 66318, 05315-970, S\~ao Paulo, SP, Brazil}
\affiliation{$^{6}$ Instituto de F\'isica, Universidade Federal do Rio de Janeiro, C.P. 68528, 21945-970, Rio de Janeiro, RJ , Brazil}
\affiliation{$^{7}$ Instituto de F\'isica, Universidade Federal Fluminense, 24210-346, Niter\'oi, RJ, Brazil}

\date{Jan. 3rd, 2019}

\begin{abstract}

Multiplicity fluctuations are one of the most crucial observables in the Beam Energy Scan program of the Relativistic Heavy Ion Collider. 
It is understood that they can be utilized to probe the whereabouts of the critical point on the phase diagram of the QCD matter.
However, a significant portion of these fluctuations is, apart from that related to the QCD phase transition, attributed to the other origins, which we refer to as ``noncritical" ones.
The present study is dedicated to the noncritical aspects of the multiplicity fluctuations in heavy-ion collisions.
In particular, we focus on those of dynamical origin, such as the hydrodynamic expansion of the system and the event-by-event initial fluctuations, in addition to the usual thermal fluctuations, finite volume corrections, and resonance decay at the freeze-out surface.
The obtained results are compared to those of the hadronic resonance gas model as well as to the experimental data.


\end{abstract}

\maketitle

\section{I. Introduction}

The ongoing Beam Energy Scan (BES) program~\cite{RHIC-star-bes-01, RHIC-star-bes-03, RHIC-star-bes-05} at the Relativistic Heavy Ion Collider (RHIC) is dedicated to exploring the phase diagram of the strongly interacting nuclear matter.
For Au+Au collisions from 3.0 to 62.4 GeV, precise measurements are being realized for the high baryon density region of the QCD matter regarding the critical endpoint of expected phase transition. 
In principle, the dynamics of such phase transitions are described by the Quantum Chromodynamics (QCD).
One intriguing characteristic of the system concerns the chiral symmetry.
Many theoretical efforts have been devoted concerning its spontaneously breaking in the QCD vacuum, as well as the restoration at the extremely hot or dense environment.
There, quarks and gluons are the relevant degrees of freedom through the deconfinement transition from the hadronic state of matter.
Lattice QCD studies~\cite{lattice-01,lattice-02} demonstrated that the transition of the system is a smooth crossover at vanishing baryon density and large strange quark mass.
At finite chemical potential, on the other hand, a variety of models~\cite{Halasz:1998qr, Berges:1998rc, Stephanov:1998dy, Schwarz:1999dj, Fodor:2004nz} predict the occurrence of a first-order phase transition between the hadronic and quark-gluon plasma (QGP), sometimes accompanied by a very complex phase structure.
These results indicate there exists a critical endpoint which is located somewhere on the QCD phase diagram where the line of first-order phase transitions terminates.
The transition is expected to be of second-order at this point.
Among other established goals, the BES program is driven by the search for the critical endpoint.
Intuitively, one might look for quantities that are sensitive to the underlying physics while accessible experimentally.
The higher cumulants of conserved charges and combinations of them, such as cumulant ratios, are candidates for such observables.
These quantities fulfill the requirement as they carry vital information on the primordial medium created in the collisions.
Moreover, it has been suggested~\cite{qcd-phase-fluctuations-review-02} that they are sensitive to the phase structure of the QCD matter, and in particular, the whereabouts of the critical point.
In this regard, recently, multiplicity fluctuations have drawn much attention as one of the key observables.

In fact, the experimentally observed multiplicity fluctuations are governed by various distinct mechanisms~\cite{qcd-phase-fluctuations-review-01,qcd-phase-fluctuations-review-02} associated with the physical system in question.
As a thermodynamical system, a considerable portion of the measured multiplicity fluctuation comes from the thermal fluctuations.
Calculations have been carried out in terms of the Hadron Resonance Gas (HRG) models in the grand canonical ensemble (GCE)~\cite{statistical-model-05,statistical-model-06,statistical-model-07} or canonical ensemble regarding conserved charges~\cite{statistical-model-03,statistical-model-04,statistical-model-08}.
For the latter, the conditions for the conservation of net-charges are explicitly considered, and the effect was shown to be substantial.
In addition, resonance decay was shown to cause nonnegligible deviation from pure statistical distributions~\cite{statistical-model-03,statistical-model-04,statistical-model-07}.
For the most part, the obtained results~\cite{statistical-model-05,statistical-model-06,statistical-model-07,statistical-model-08,statistical-model-09,statistical-model-10} are manifestly consistent with the experimental data~\cite{RHIC-star-mul-fluctuations-01,RHIC-star-mul-fluctuations-02}.
On the other hand, various physical quantities become divergent, such as correlation length and particle fluctuations, as the system approaches the critical point of a system in thermal equilibrium.
While a quantitative description of the critical phenomena is provided by the theory of renormalization group, owing to the sophistication of the problem at hand, one usually resorts to phenomenological approaches, such as the $\sigma$ model~\cite{qcd-phase-fluctuations-01}.
It has been speculated~\cite{qcd-phase-fluctuations-02,qcd-phase-fluctuations-03,qcd-phase-fluctuations-04} that the normalized fourth order cumulant of multiplicity distribution might be a non-monotonic function of collision energy. 
In reality, instead of being stationary, homogeneous, and infinite in extension, the system created in heavy-ion collisions evolves rapidly in time, it is highly inhomogeneous while occupies only a small volume in space.
Meanwhile, the measurements are carried out on the freeze-out surface in terms of hadronized particles, which might be not so close to the critical point in the phase diagram. 
In this regard, the effect of the critical endpoint on the dynamics of the system is essential.
Such tentatives~\cite{hydro-chiral-01} eventually leads to a variety of models.
For instance, the chiral fluid dynamics~\cite{hydro-chiral-04,hydro-chiral-07,hydro-chiral-08,hydro-chiral-09} treats quarks as an equilibrated heat bath. Subsequently, a Langevin equation is obtained for the chiral field.
On the other hand, Hydro+~\cite{hydro-chiral-sigma-01} approach focuses on the critical slowing-down when the time scale to achieve local equilibrium becomes comparable to that for global equilibrium.
Moreover, even in the framework of conventional hydrodynamics, the existence of a critical point may impact the temporal evolution via its modification to the equation of state (EoS).
Also, there are additional sources which may affect the resulting multiplicity fluctuations.
To be more specific, thermal~\cite{hydro-fluctuations-03} and non-equilibrium~\cite{hydro-fluctuations-02} fluctuations on freeze-out surface, experimental uncertainties and cuts, and other spurious contributions may substantially attenuate the measured signals~\cite{qcd-phase-fluctuations-08,qcd-phase-fluctuations-09}.

In the present work, we focus on a hydrodynamic study of the multiplicity fluctuations, which is mainly based on the scenario of HRG models.
Our approach takes into consideration thermal fluctuations by using the formalism of GCE.
Also, volume correction, as well as resonance decay, are considered regarding hadron emission.
The hydrodynamic evolution is expressed in terms of the Smoothed Particle Hydrodynamics (SPH) algorithm.
In our model, every elementary degree of freedom of the system, namely, a small fluid element denoted by an SPH particle, is treated as a quantum GCE.
In comparison with statistical model approaches, system expansion is encoded in terms of freeze-out surface.
As a result, the resultant element of the freeze-out surface may also possess nonvanishing spatial component.
Moreover, event-by-event initial conditions (IC) are explicitly considered and shown to play a significant role in the resulting quantities.

The paper is organized as follows.
In the following section, we briefly review relevant aspects concerning thermodynamical fluctuations and resonance decay.
We give an account of the specific implementation for the hydrodynamic code SPheRIO in Section III.
Numerical simulations are carried out, and the results are presented and discussed in Section IV.
The last section is dedicated to concluding remarks.

\section{II. Thermodynamical fluctuations and resonance decay}

For a static ideal gas, the particle number fluctuations can be measured regarding the variance and covariance of particle numbers.
These quantities can be readily evaluated by quantum statistical physics~\cite{book-landau-5}.
To be specific, the GCE average value and variance of the occupation density in the momentum space read~\cite{statistical-model-03,statistical-model-04}
\begin{eqnarray}
\langle n_{p,i} \rangle= \frac{1}{\exp\left[(\sqrt{p^2+m_i^2}-\mu_i)/T\right]-\gamma_i} , 
\end{eqnarray}
\begin{eqnarray}
\langle \Delta n_{p,i}^2 \rangle \equiv  \langle (n_{p,i} - \langle n_{p,i} \rangle)^2 \rangle = \langle n_{p,i} \rangle (1+\gamma_i \langle n_{p,i} \rangle) ,\label{Delta_npi2}
\end{eqnarray}
where $p$ is the momentum, the subscript $i$ indicates particle species, $T$ is the temperature, $m_i$ and $\mu_i$ are the particle mass and chemical potential respectively, $\gamma_i$ corresponds to Bose (+1), Fermi (-1) or Boltzmann (0) statistics.

For systems at chemical equilibrium, one has
\begin{eqnarray}
\mu_i=q_i\mu_Q +b_i\mu_B+s_i\mu_S ,
\end{eqnarray}
where $q_i, b_i, s_i$ are the electric charge, baryon number and strangeness of particle species $i$, and $\mu_Q, \mu_B, \mu_S$ are the chemical potentials of the corresponding conserved charges.

In our present approach, the fluctuations are independent for different particle species as well as different momentum space, the covariance is found to be
\begin{eqnarray}
\langle \Delta n_{p,i} \Delta n_{k,j} \rangle = \delta_{ij}\delta_{pk}v_{p,i}^2 , \label{covnpi}
\end{eqnarray}
where $\Delta n_{p,i} = n_{p,i} - \langle n_{p,i}\rangle$, and $v_{p,i}^2=\langle \Delta n_{p,i}^2 \rangle$, given in Eq.~(\ref{Delta_npi2}).

By summing up different momentum states, the average number of particles of species $i$ is given by
\begin{eqnarray}
\langle N_{i} \rangle= \sum_p \langle n_{p,i} \rangle = \frac{g_i V}{2\pi^2}\int_0^{\infty} p^2 dp \langle n_{p,i} \rangle . \label{avgNi}
\end{eqnarray}

The variance $\sigma^2$ for species $i$ reads
\begin{eqnarray}
 \langle\left(\Delta N_i\right)^2\rangle=T\left(\frac{\partial N_i}{\partial \mu}\right)_T , \label{thermoVariance}
\end{eqnarray}
and similarly, since the covariance between different particle species vanishes, we have
\begin{eqnarray}
\langle \Delta N_{i}\Delta N_{j} \rangle= \sum_{p,k} \langle \Delta n_{p,i}\Delta n_{k,j} \rangle = \delta_{ij}\sum_p v_{p,i}^2 .\label{thermoCovariance}
\end{eqnarray}

Besides, higher statistical moments of multiplicity distributions like skewness $S\propto \langle \Delta N^3\rangle$ and kurtosis $\kappa \propto \langle \Delta N^4\rangle$ are also of particular importance.
These quantities are sensitive enough to the correlation length. 
Furthermore, products $\kappa\sigma^2$ and $S\sigma$ are directly related to the ratios of the cumulants of particle numbers.
For a homogeneous system, these quantities are same the ratios of susceptibilities where the volume and temperature-dependent terms cancel out~\cite{qcd-phase-fluctuations-review-02}.
While such higher moments can be evaluated similarly, the calculations, as well as the resulting expressions, are somewhat tedious.
Therefore, we delegate a succinct account for the relevant expressions to the Appendix of the present paper.

In order to consider the effect of conserved charges, one may follow Refs.~\cite{statistical-model-03,statistical-model-04} to insert some additional factor into the phase space integral of the grand partition function.
To be specific, 
\begin{eqnarray}
\prod\limits_{i}\frac{1}{2\pi} \int_0^{2\pi} d\phi_i\exp\left[-i Q_i \phi_i\right] ,
\end{eqnarray}
where $Q_i$ stands for the total charge of type $i$, for instance, $Q_i = Q, B, S$, etc.
The integral can be evaluated by further making use of the saddle point expansion technique, and therefore approximated but analytic results can be obtained.
The resulting partition function is usually referred to as ``canonical" in the literature.
We note that for the above prescription, the conservation is demanded for specific net-charges but not for individual particle species.
Otherwise, the variance of any particle species shall vanish by definition.
It was shown~\cite{statistical-model-03} that, depending on specific model parameters, the effect of conserved charges could be substantial.

The resonance decay can be considered by introducing the following generating function~\cite{statistical-model-03}
\begin{eqnarray}
G\equiv \prod_R \left(\sum_r b_r^R \prod_i\lambda_i^{n_{i,r}^R}\right)^{N_R} , \label{resonanceDecayGenerator}
\end{eqnarray}
where for a given resonance $R$, a specific decay channel is denoted by $r$ with the branching ratio $b_r^R$.
Also, ${n_{i,r}^R}$ indicates the number of particles $i$ obtained through the decay channel $r$ of the resonance in question.
Here $\lambda_i$ is the ``external source" which will be taken to be $1$ by the end of the calculations.
The resulting particle number of a specific particle species $i$ can be obtained by the operation $\lambda_i\frac{\partial }{\partial \lambda_i}$.
As a result, one finds
\begin{eqnarray}
\overline{N}_i\equiv \sum_R\langle N_i\rangle =\lambda_i\frac{\partial }{\partial \lambda_i}G= \sum_R N_R \sum_r b_r^R n_{i,r}^R\equiv \sum_R{N_R}\langle n_i\rangle_R ,
\end{eqnarray}
\begin{equation} 
\begin{aligned}
&\overline{N_i N_j} \equiv \sum_R\langle N_i N_j\rangle_R+\sum_{R\ne R'}\langle N_i N_j \rangle_{R,R'} =\lambda_i\frac{\partial }{\partial \lambda_i}\left(\lambda_j\frac{\partial }{\partial \lambda_j}G\right)  \\
=&\sum_R\left[N_R(N_R-1)\langle n_i\rangle_R\langle n_j\rangle_R+N_R\langle n_i n_j\rangle_R \right]+\sum_{R\ne R'} N_R N_{R'}\langle n_i\rangle_R\langle n_j\rangle_{R'} . \label{res2}
\end{aligned}
\end{equation}
Here we have used an overline ``$\overline{\quad}$" to indicate the resulting ensemble average value after considering all possible decay modes.
Whereas, $\langle \cdots\rangle_R$ means the average over different decay modes for a given resonance $R$.
For instance, $\langle n_i n_j\rangle_R = \sum_r b_r^R n_{i,r}^Rn_{j,r}^R$.
The overlined value is thus obtained by summing up all the contributions from different resonances.
The derivation for other relevant higher moments used in this work can be found in the Appendix.

When one evaluates the variance and covariance, which involves more than one particle, it is noted that the contribution may come from a variety of possible decay processes.
For instance, two decayed particles might originate from the same resonance, two distinct resonances of the same type, and two different resonance.
However, all these possibilities are automatically taken care of as referred from the last line of Eq.~(\ref{res2}).

Subsequently, one may proceed to evaluate experimental observables.
One such quantity frequently cited in the literature is the scaled variance. 
For a given initial resonance distribution, it is found to be
\begin{eqnarray}
\omega_R^{i*}\equiv \frac{\langle N_i^2\rangle_R-\langle N_i\rangle_R^2}{\langle N_i\rangle_R}=\frac{\langle n_i^2\rangle_R-\langle n_i\rangle_R^2}{\langle n_i\rangle_R}=\frac{\sum_r b_r^R\left(n_{i,r}^R\right)^2-\left(\sum_r b_r^R n_{i,r}^R\right)^2}{\sum_r b_r^R n_{i,r}^R} .
\end{eqnarray}
The resulting expression taking into account for all different resonances reads
\begin{eqnarray}
\overline{\omega^{i*}_R}= \frac{\overline{N_i^2}-\overline{ N_i}^2}{\overline{ N_i}}=\frac{\sum_R N_R\langle n_i^2\rangle_R-\sum_R N_R\langle n_i\rangle_R^2}{\sum_R N_R\langle n_i\rangle_R} .
\end{eqnarray}

In realistic events, resonance yields $N_R$ also fluctuate, and the resultant scaled variance reads
\begin{eqnarray}
\omega^{i*}\equiv \frac{\langle\overline{N_i^2}\rangle_T-\langle\overline{ N_i}^2\rangle_T}{\langle\overline{ N_i}\rangle_T}=\overline{\omega^{i*}_R} + \sum_R\langle n_i\rangle_R \omega_R ,
\end{eqnarray}
where
\begin{eqnarray}
\omega_R \equiv \frac{\langle {N_R^2}\rangle_T-\langle { N_R}\rangle^2_T}{\langle {N_R}\rangle_T} 
\end{eqnarray}
is the scaled variance of the resonance $R$.

If the system is static and in thermal and chemical equilibrium, the thermal fluctuations used in the above expressions are those discussed above in Eqs.~(\ref{thermoVariance}-\ref{thermoCovariance}).

\section{III. A hydrodynamic approach}

In this section, we elaborate an approach which incorporates the effect of hydrodynamical evolution of the system, together with the event-by-event fluctuating IC on multiplicity fluctuations.
To take into consideration the temporal expansion into our framework, we employ SPheRIO~\cite{sph-review-01}, a hydrodynamic code for an ideal relativistic fluid based on SPH algorithm. 
In this approach, the fluid motion is represented in terms of discrete Lagrangian coordinates, known as SPH particle. 
In the case of an ideal fluid, the latter is assigned with a given fraction of conserved quantities, say, the entropy and also the baryon number.
In term of the SPH particle degree of freedom, the equation of motion can be derived by using the variational principle. 
We neglect in the present work any dissipative effects and assume Cooper-Frye sudden freeze-out take place at constant temperatures.
The latter, when transformed into the local rest frame, provides the baseline to evaluate the thermal fluctuations at the moment of hadronization. 
We do not introduce any additional free parameter into the model as the existing ones that have been determined as to appropriately reproduce the experimental data regarding the particle spectra~\cite{sph-eos-02,sph-vn-04,sph-v2-02,sph-corr-ev-04,sph-eos-03,sph-vn-04,sph-corr-ev-06,sph-corr-ev-08,sph-corr-ev-09,sph-vn-06}.

On the freeze-out surface, every small fluid element, that is, in our case, an SPH particle is treated as a GCE for a given temperature and the mean baryon number.
One might proceed further to take into account conserved charges, as discussed in the previous section.
Unfortunately, the latter is highly nontrivial, owing to precisely the same difficulties to explicitly incorporate global charge conservation at hadronization in most hydrodynamical models. 
A hydrodynamic event is a collection of GCE ensembles represented by SPH particles.
While in the fluid dynamical representation, it naturally gives the correct value for the total charge of the system on average, once we introduce the freeze-out for hadronization via GCE, the exact charge conservation becomes extremely difficult to be implemented numerically.
To be specific, this is because the momentum space integral involving a conserved total charge is then to be carried out on all individual freeze-out surface elements resolved numerically.
It is noted that significant progress has been achieved recently about implementing canonical or microcanonical systems on the freeze-out surface~\cite{hydro-fluctuations-04}.
As a first approximation, however, we will ignore the condition of charge conservation in our present approach.

For each fluid element at the moment of hadronization, it is in local equilibrium.
In this case, however, the volume in Eq.~(\ref{avgNi}) becomes anisotropic.
It should be replaced by a time-like 3-surface.
Moreover, the particle number flux also depends on the frame of reference, and integral in momentum space should be modified accordingly.
To be specific, the average number of particles of species $i$ is replaced by the following covariant form
\begin{eqnarray}
\frac{E d^3N_i}{d p^3} = \frac{d^3N_i}{2\pi p_T dp_T dy} = \int_\sigma d\sigma_\mu p^\mu \langle n_{i}(u, p, x) \rangle ,\label{d3n-hydro}
\end{eqnarray}
which is expressed in terms of dynamical variables such as rapidity $y$ and transverse momentum $p_T$.
As mentioned before, the volume has been substituted by an integral carried out on $\sigma_\mu$, an element determined by the hydrodynamical calculations.
For the latter, if only its time-component is non-vanishing, contracting with $p^\mu$ and integrating in momentum will bring it right back to Eq.~(\ref{avgNi}), since
\begin{eqnarray}
\langle n_{i}(u, p, x) \rangle \equiv \langle n_{i}(u\cdot p, x) \rangle = \frac{1}{\exp\left[(u(x)\cdot p-\mu_i(x))/T\right]-\gamma_i}
\end{eqnarray}
is the local occupation density in the co-moving frame.

Following the spirit of the SPH method, Eq.~(\ref{d3n-hydro}) can be rewritten in terms of SPH degrees of freedom.
One finds,
\begin{eqnarray}
E\frac{d^3N_i}{dp^3}=\sum_j \frac{\nu_j n_{j\mu}p^{\mu}}{s_j|n_{j\rho}u_j^{\rho}|}\theta(u_{j\delta}p^{\delta})\langle n_{i}(u_{j\nu}p^{\nu}, x) \rangle, \label{d3n-sph}
\end{eqnarray}
where the sum in $j$ is carried out for SPH particles, $\nu_j$ and $s_j$ denote the total entropy and entropy density of the $j$-th SPH particle.
Therefore, the ensemble average of particle number reads
\begin{eqnarray}
\langle N_{i} \rangle=\int p_{\bot}dp_{\bot}dy d\phi \sum_j \frac{\nu_j n_{j\mu}p^{\mu}}{s_j|n_{j\rho}u_j^{\rho}|}\theta(u_{j\delta}p^{\delta})\langle n_{i}(u_{j\nu}p^{\nu}, x) \rangle ,\label{NavgSPH}
\end{eqnarray}
where, again, different fluid elements are treated to be statistically independent, as they are individual GCEs.

We note that the Cooper-Frye formalism may lead to a negative contribution to particle flux which is stripped away by the $\theta$-function in Eq.~(\ref{d3n-hydro}).
This is a known problem which leads to a sudden increase in energy after the hadronization process.
For event-by-event fluctuating IC, the total energy discrepancy can be about 20 - 25\%, and for smoothed IC, the situation is less severe, and the amount is less than 10\%.
Similarly, the deviation of the baryon number and other conserved charges can be determined accordingly regarding the specific EoS in question.
As discussed below, the above issue regarding conservations of energy and other conserved charges might be improved by adopting a more subtle scheme of hadronization.

It is not difficult to further show that the covariance is
\begin{eqnarray}
\langle \Delta N_{i}\Delta N_{j} \rangle= \int p_{\bot}dp_{\bot}dy d\phi \sum_j \frac{\nu_j n_{j\mu}p^{\mu}}{s_j|n_{j\rho}u_j^{\rho}|}\theta(u_{j\delta}p^{\delta}) v_{i}^2(u_{j\nu}p^{\nu}, x) ,\label{NcovSPH}
\end{eqnarray}
where $v_{i}^2(u_{j\nu}p^{\nu}, x)$ follows the defintion introduced in Eq.~(\ref{covnpi}).

Eqs.~(\ref{NavgSPH}-\ref{NcovSPH}) can be readily employed to evaluate the moments and replace those for static system, for instance, Eqs.~(\ref{avgNi}), (\ref{thermoVariance}), and (\ref{thermoCovariance}).
Also, we relegate the expressions for higher moments to the Appendix.

\section{IV. Numerical results and Discussions}

\begin{figure}[htb]
\begin{tabular}{cc}
\begin{minipage}{170pt}
\centerline{\includegraphics[width=200pt]{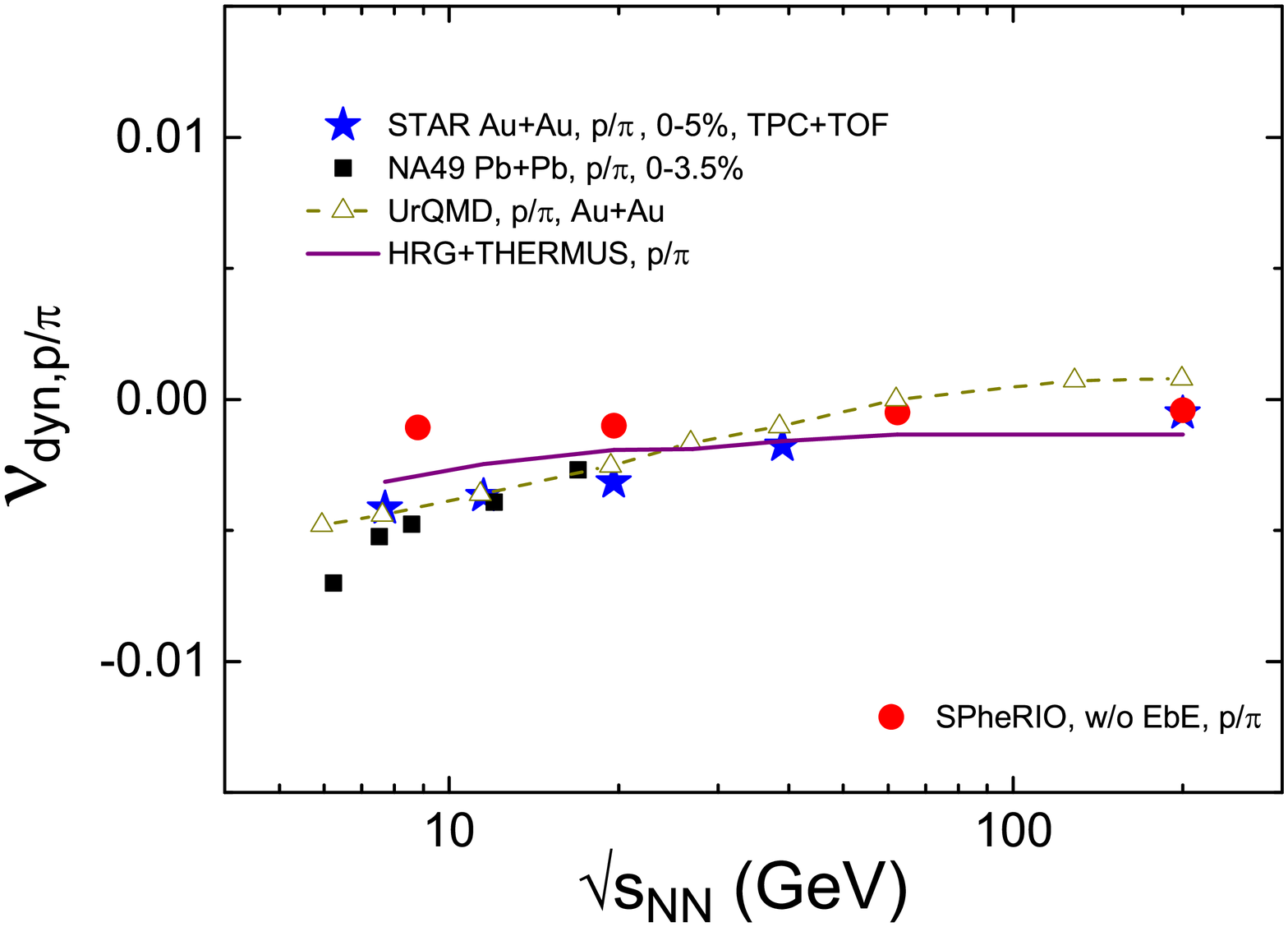}}
\end{minipage}
&
\begin{minipage}{170pt}
\centerline{\includegraphics[width=200pt]{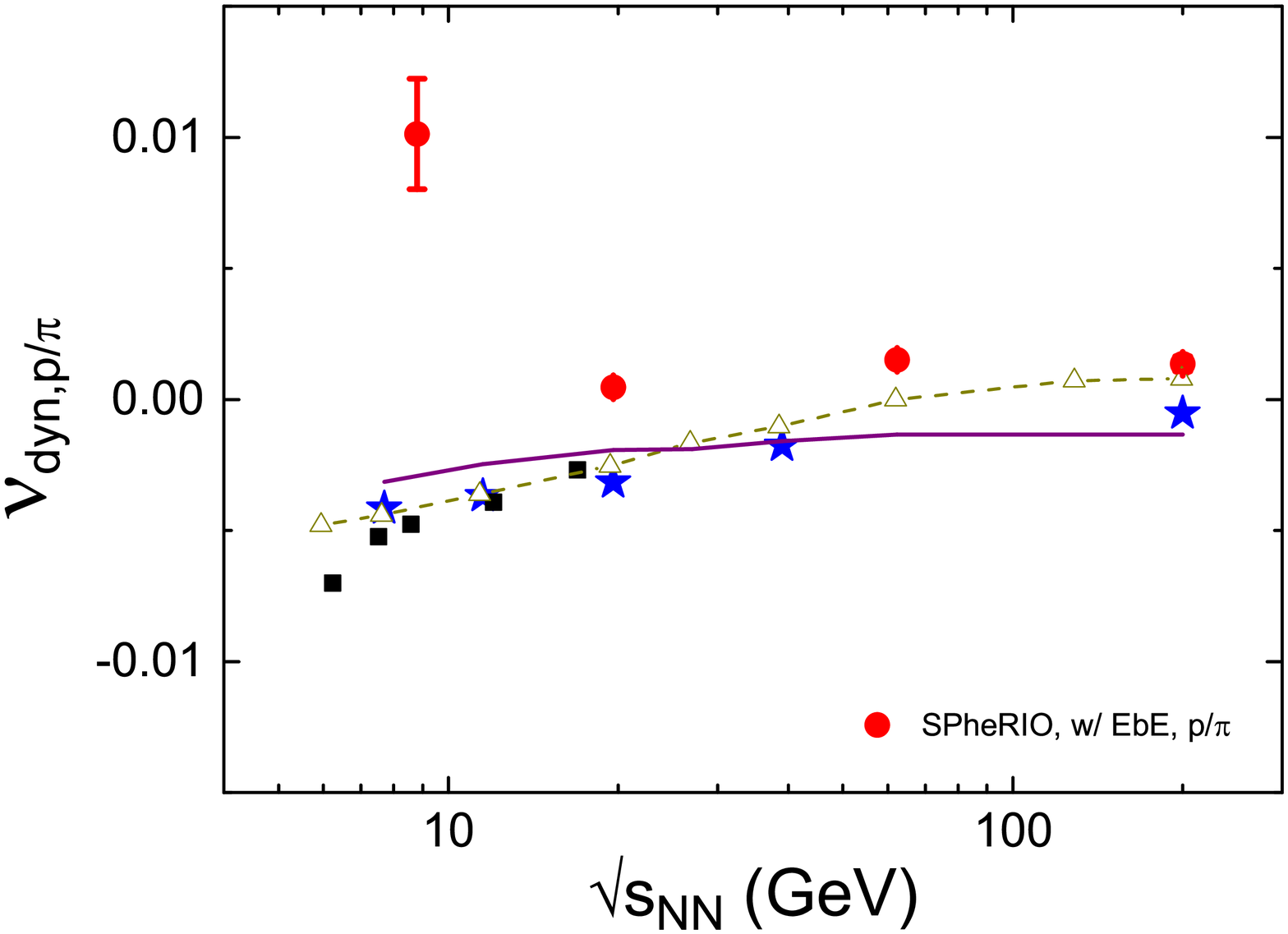}}
\end{minipage}
\\
\begin{minipage}{170pt}
\centerline{\includegraphics[width=200pt]{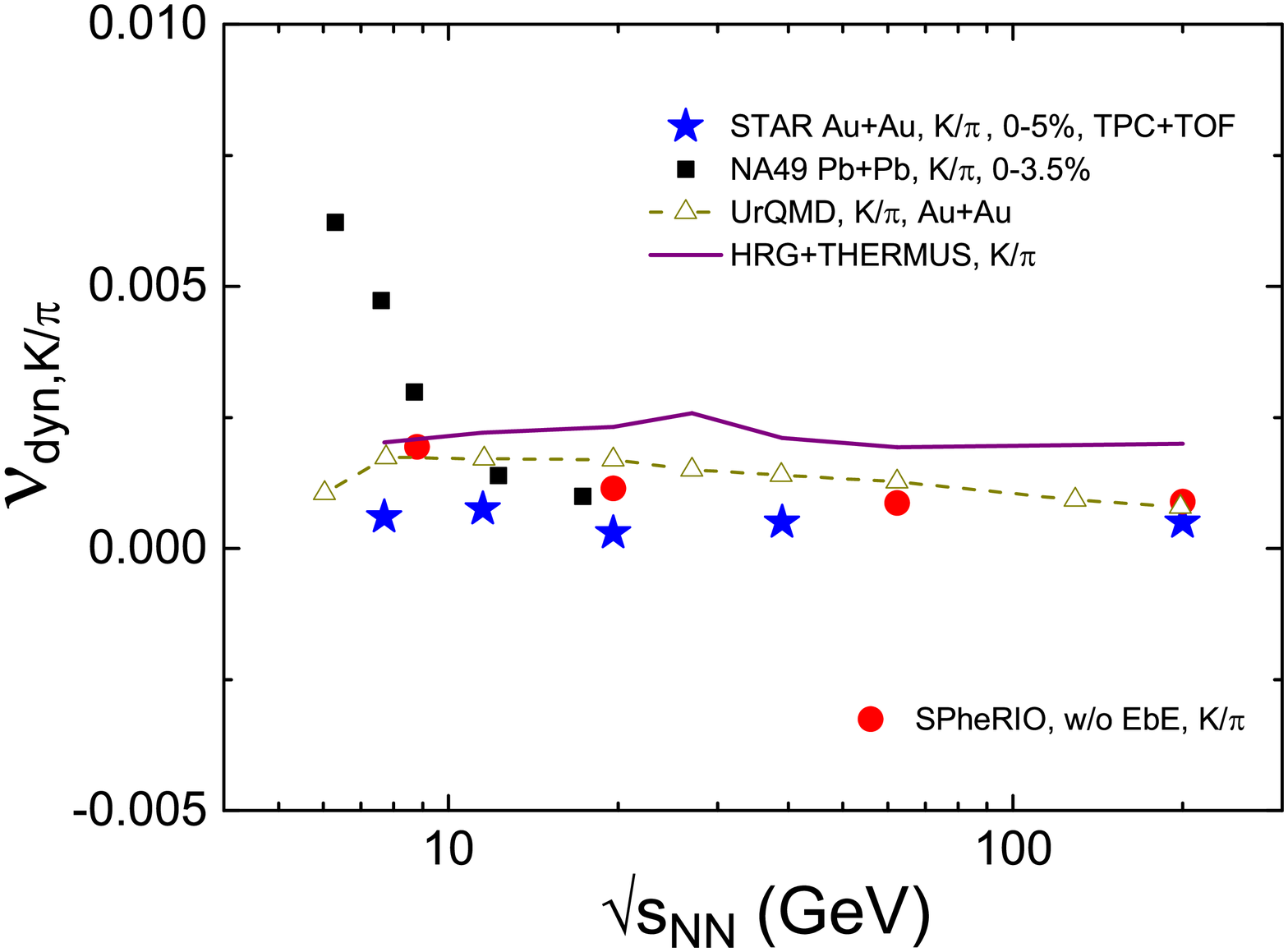}}
\end{minipage}
&
\begin{minipage}{170pt}
\centerline{\includegraphics[width=200pt]{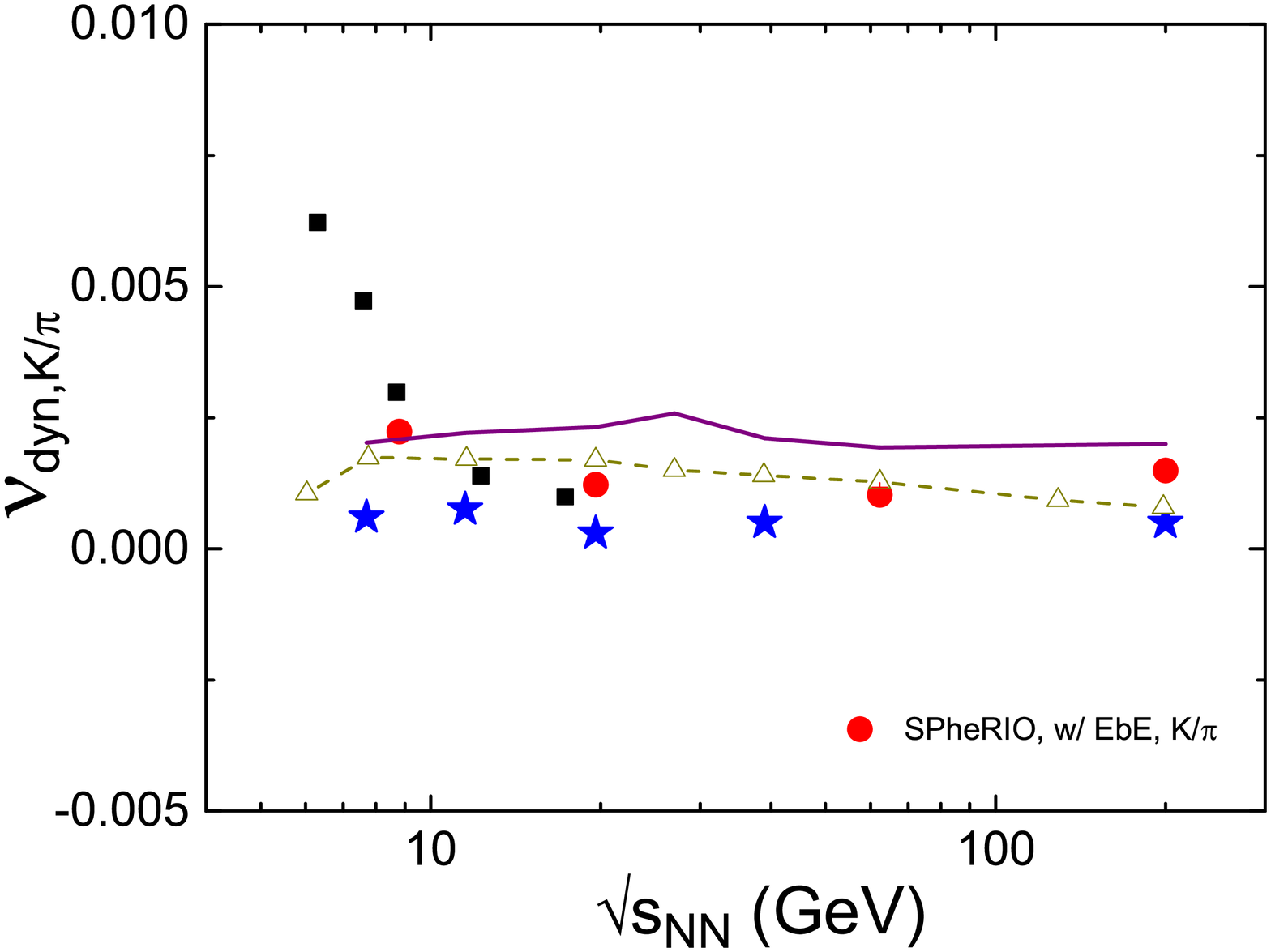}}
\end{minipage}
\\
\begin{minipage}{170pt}
\centerline{\includegraphics[width=200pt]{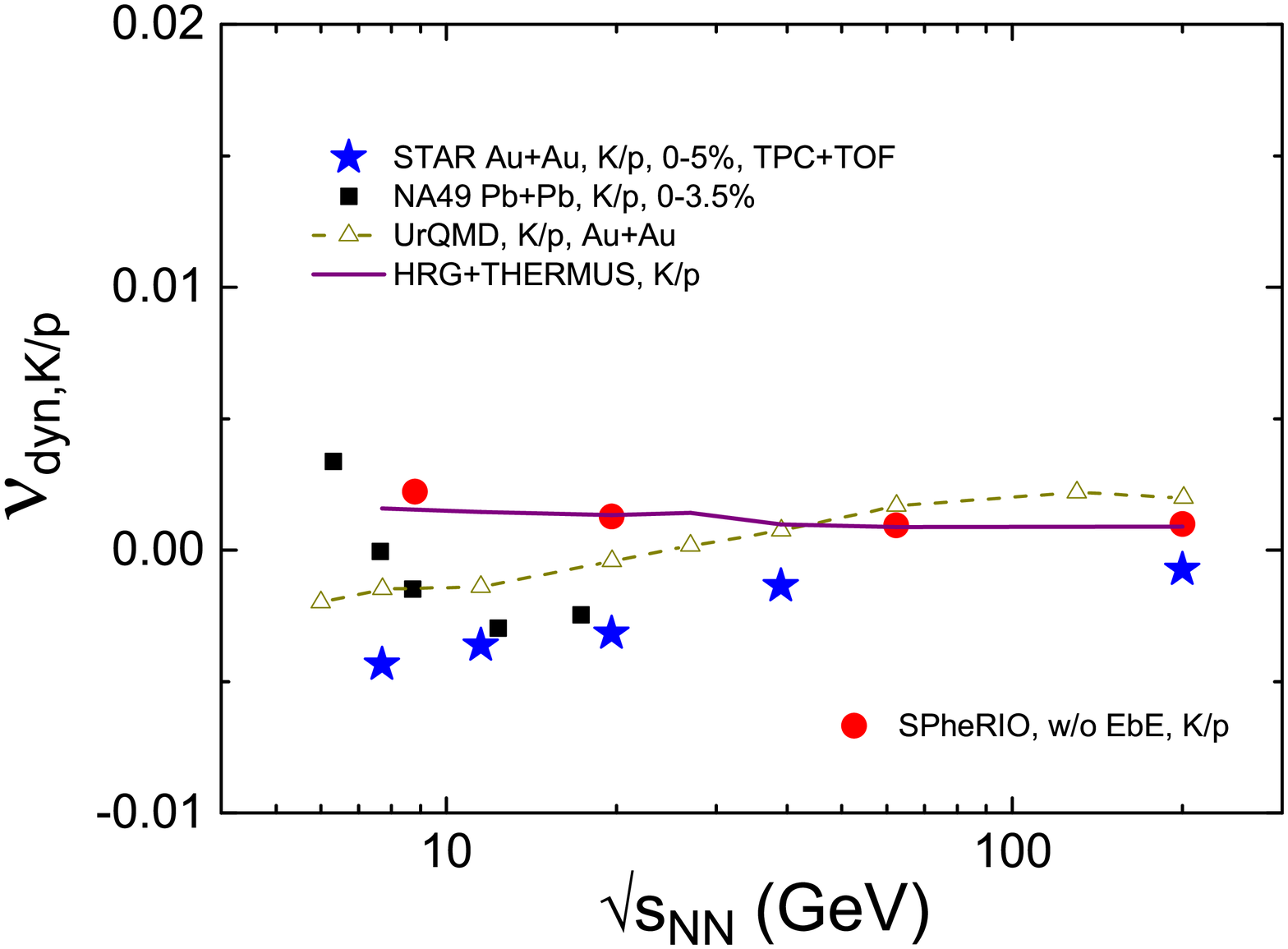}}
\end{minipage}
&
\begin{minipage}{170pt}
\centerline{\includegraphics[width=200pt]{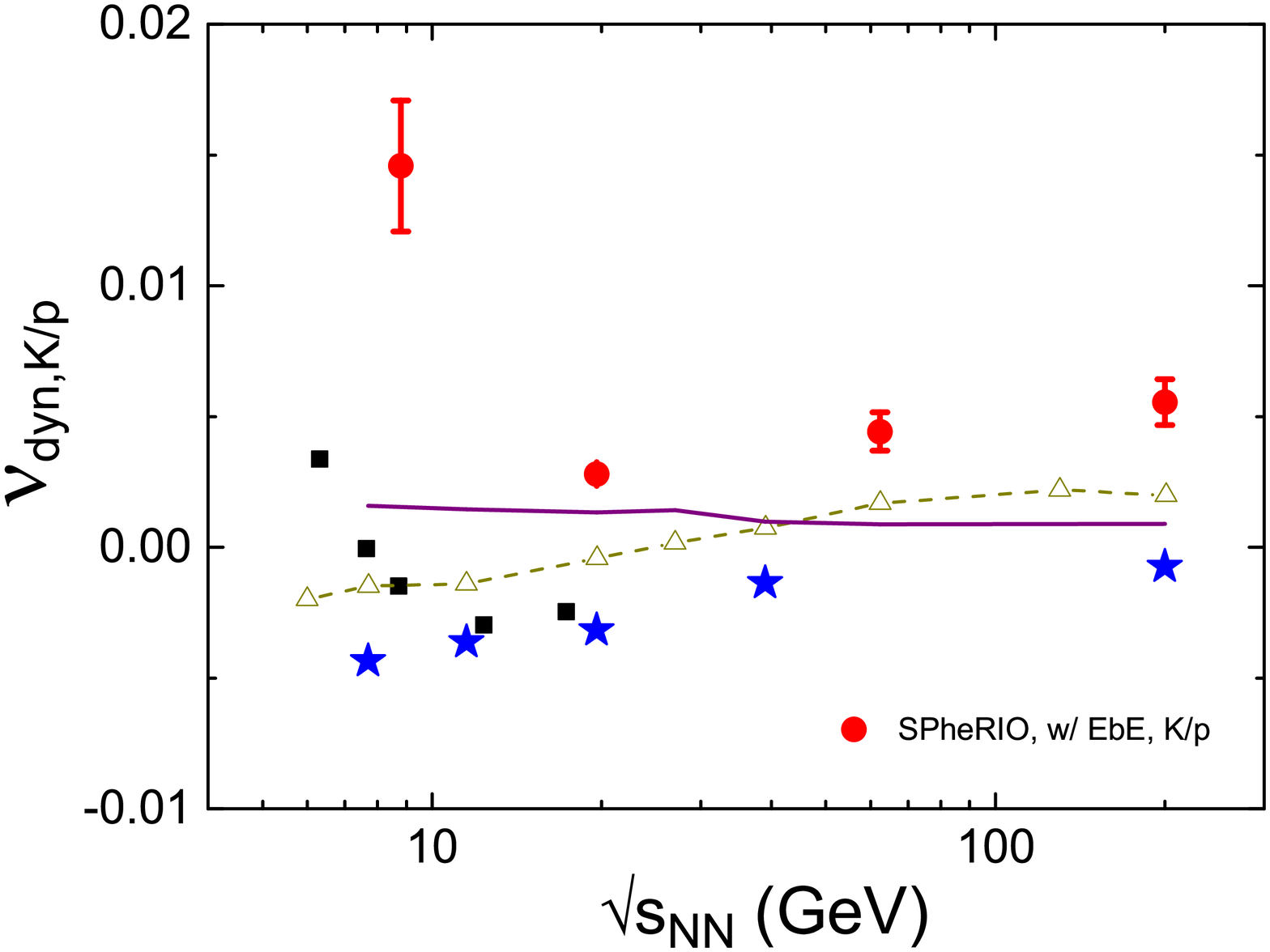}}
\end{minipage}
\end{tabular}
\caption{(Color online) The calculated dynamical fluctuations of particle ratios $p/\pi$, $K/\pi$, and $K/p$ in comparison with the data for different energies.
The experimental data are from the NA49~\cite{LHC-na49-mul-fluctuations-01} and STAR~\cite{RHIC-star-mul-fluctuations-01} collaborations.
The STAR data are for 0 - 5\% Au+Au collisions at various energies from $\sqrt{s_{NN}}=7.7$ to $200$ GeV, presented by filled blue stars.
The NA49 data are for 0 - 3.5\% Pb+Pb collisions at energies from $\sqrt{s_{NN}}=6.3$ to $17.3$ GeV, shown in filled black squares.
The SPheRIO results are given by filled red squares, for both average (left column) and event-by-event fluctuating (right column) ICs.
The UrQMD model calculations are shown in open dark-yellow triangles with dashed curves.
The HRG calculations are presented in purple solid curves.}
\label{fpratios}
\end{figure}

\begin{figure}[htb]
\begin{tabular}{cc}
\begin{minipage}{170pt}
\centerline{\includegraphics[width=200pt]{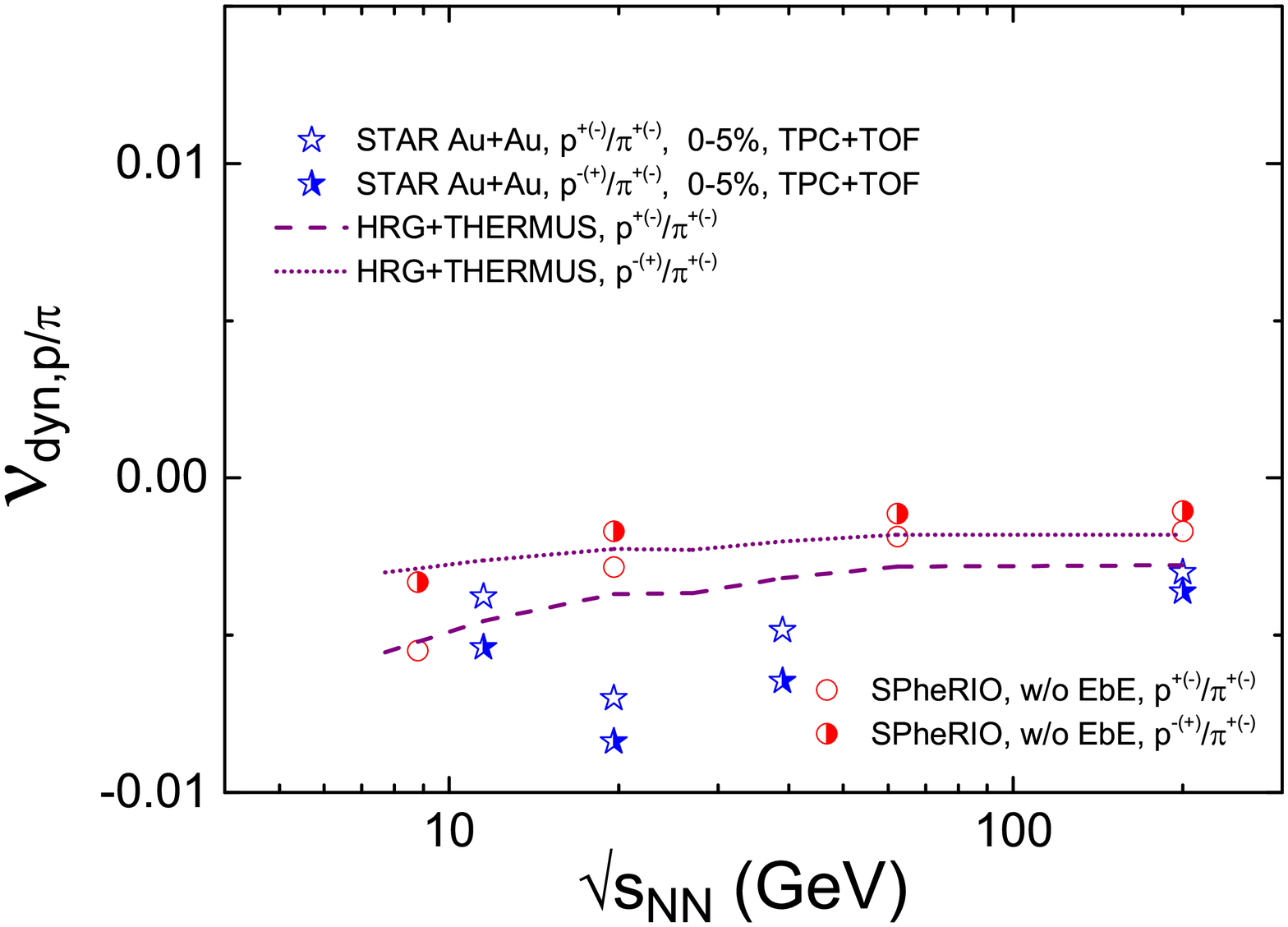}}
\end{minipage}
&
\begin{minipage}{170pt}
\centerline{\includegraphics[width=200pt]{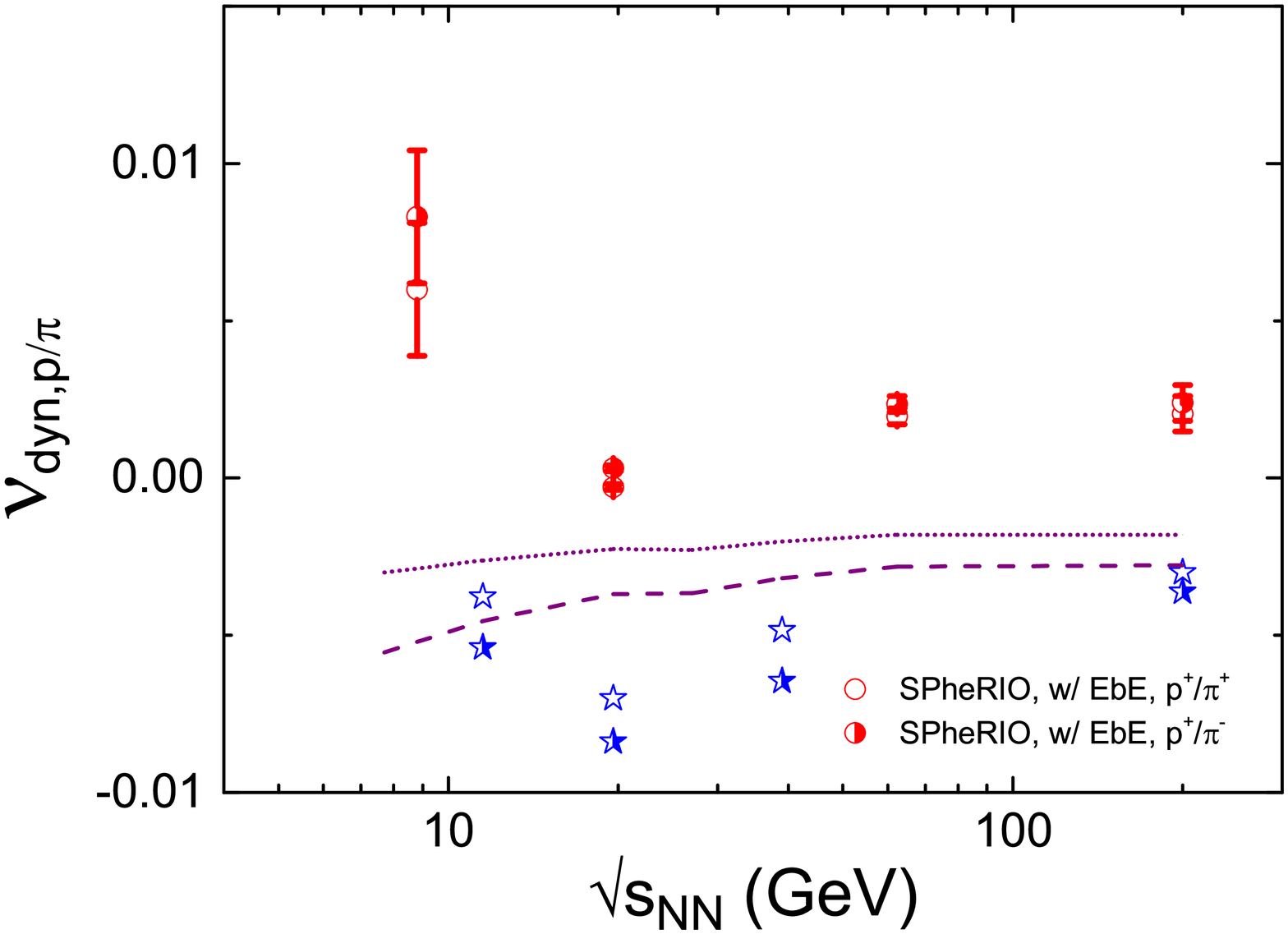}}
\end{minipage}
\\
\begin{minipage}{170pt}
\centerline{\includegraphics[width=200pt]{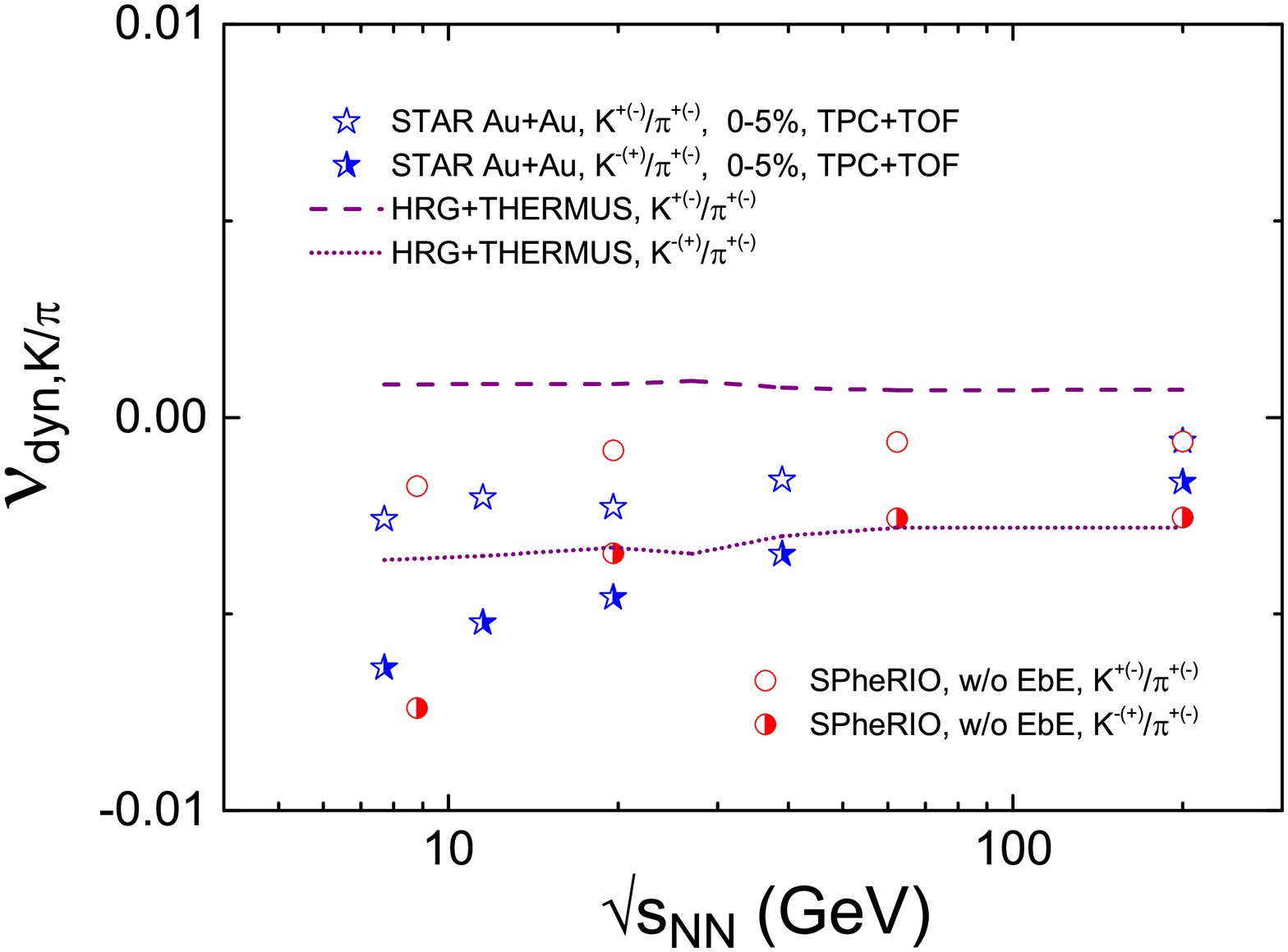}}
\end{minipage}
&
\begin{minipage}{170pt}
\centerline{\includegraphics[width=200pt]{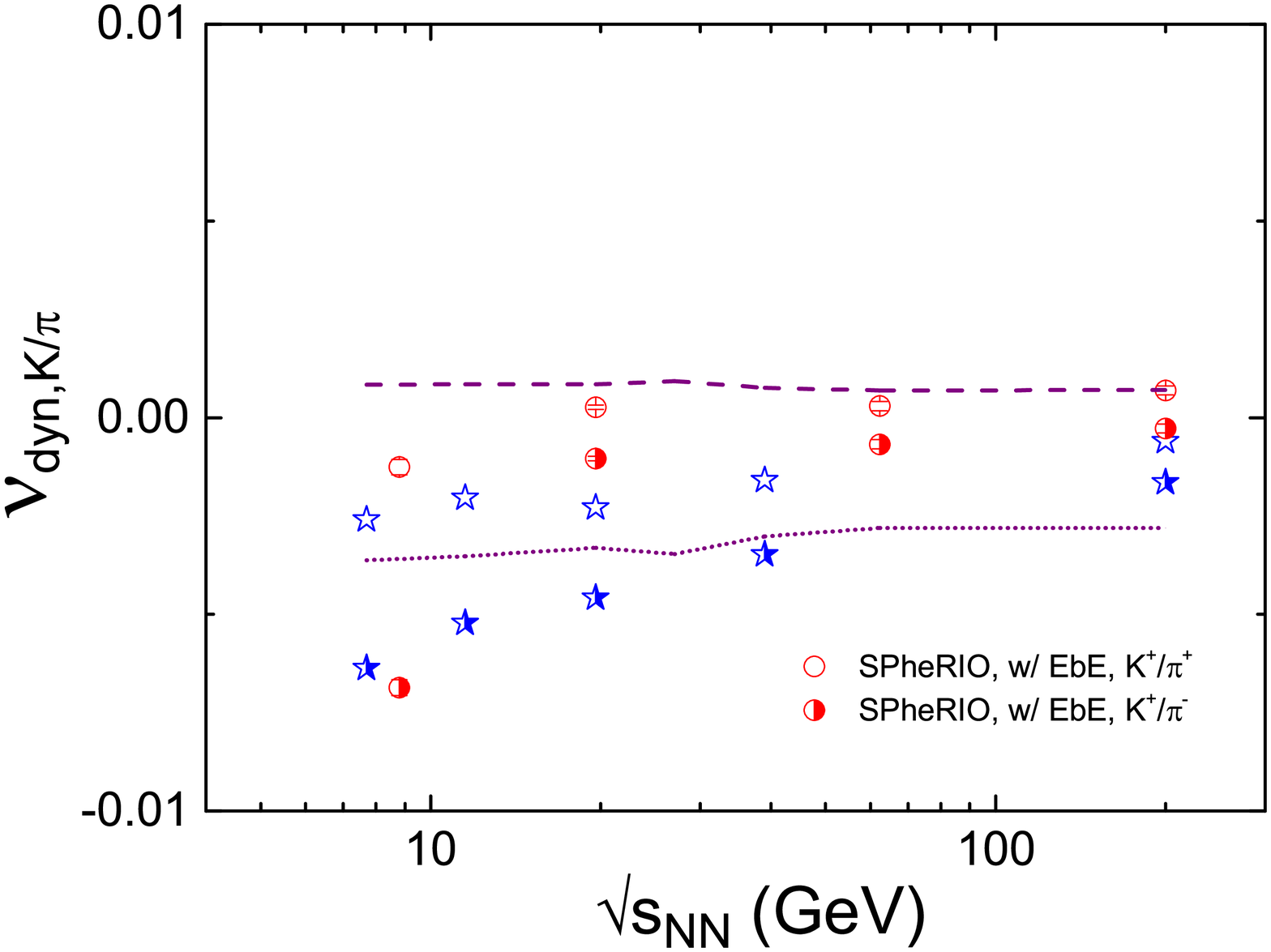}}
\end{minipage}
\\
\begin{minipage}{170pt}
\centerline{\includegraphics[width=200pt]{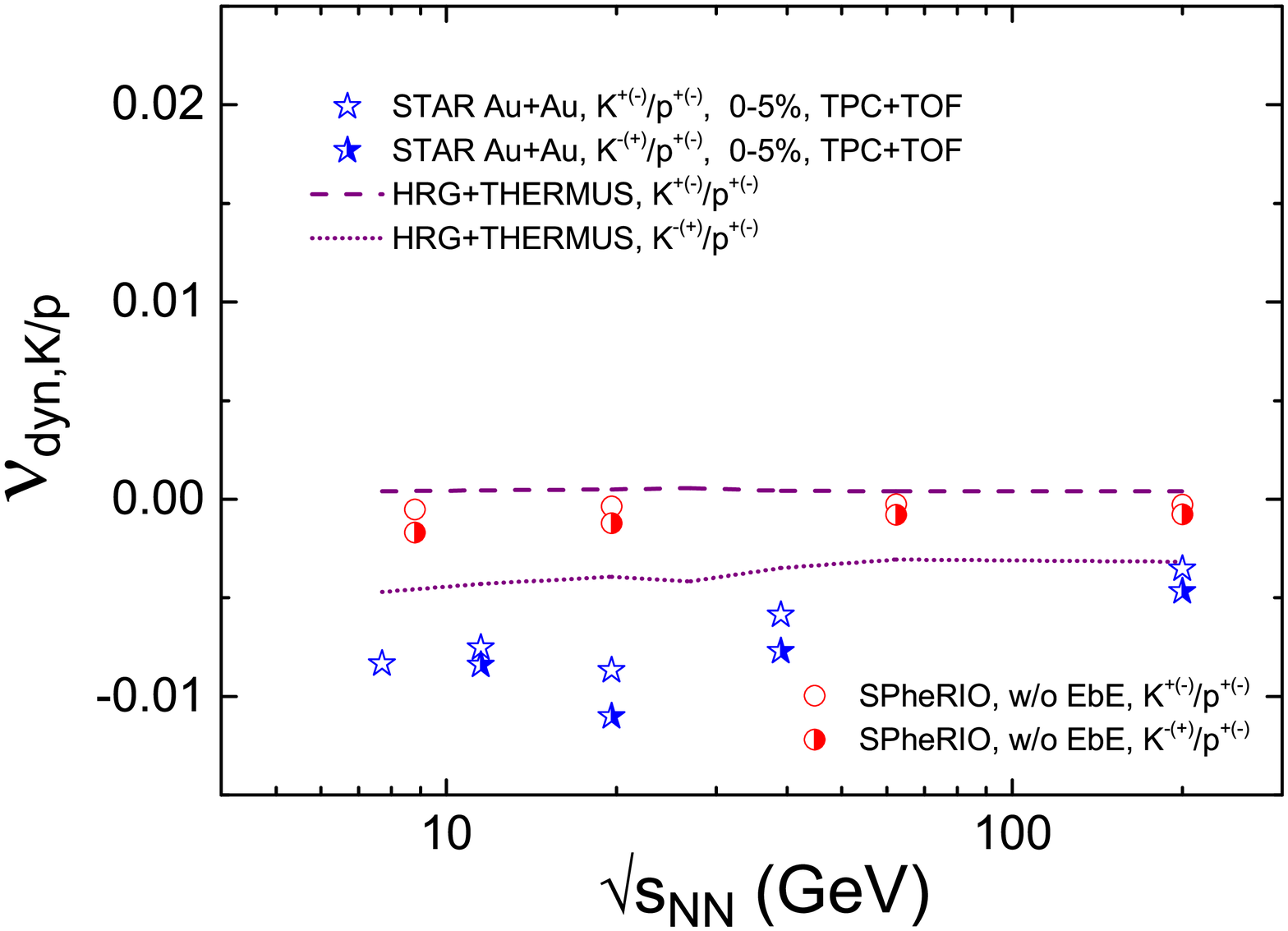}}
\end{minipage}
&
\begin{minipage}{170pt}
\centerline{\includegraphics[width=200pt]{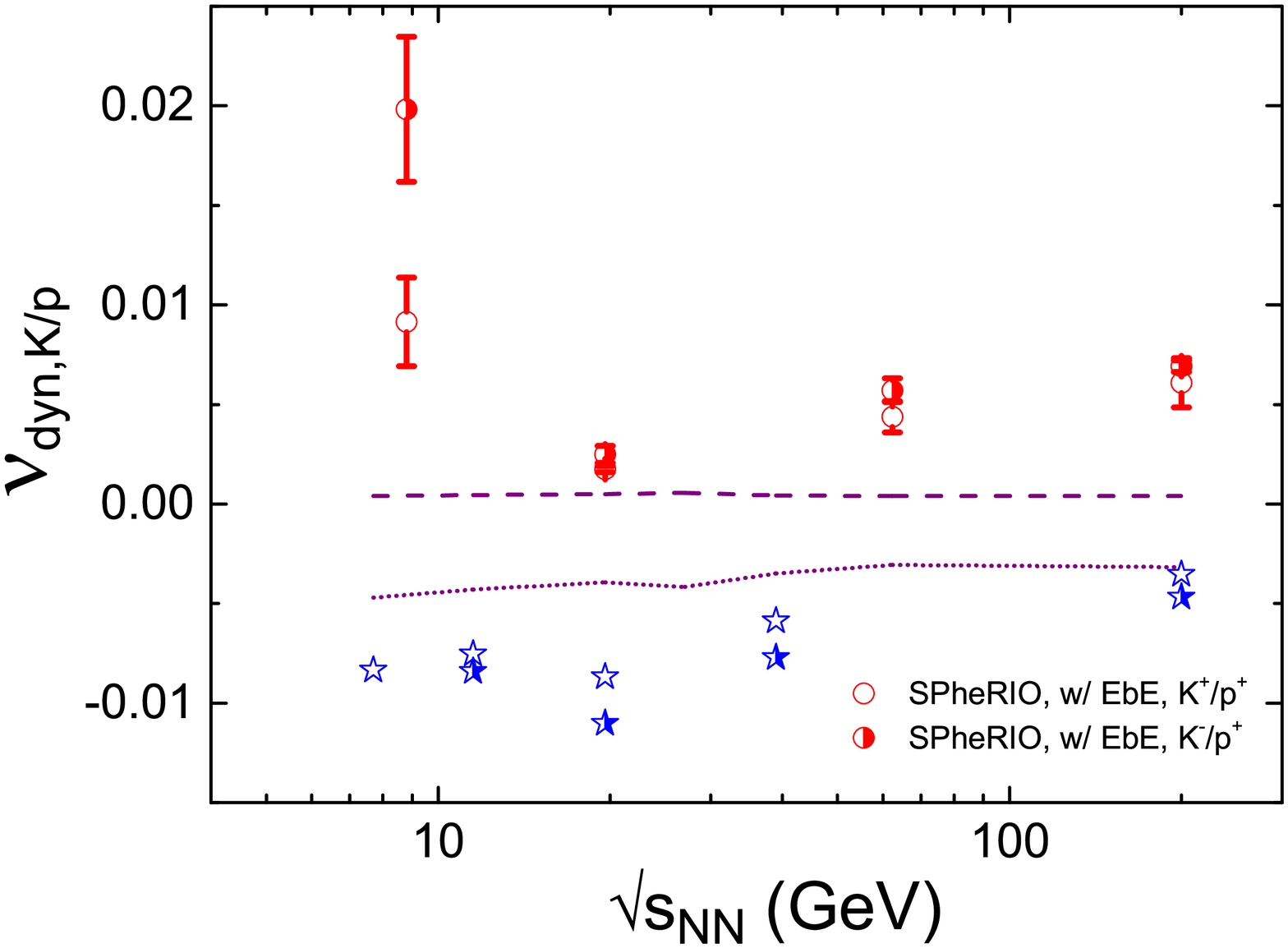}}
\end{minipage}
\end{tabular}
\caption{(Color online) The calculated dynamical fluctuations of particle ratios $p^{\pm(\mp)}/\pi^{\pm}$, $K^{\pm(\mp)}/\pi^{\pm}$, and $K^{\pm(\mp)}/p^{\pm}$ in comparison with the data for different energies.
As in Fig.~\ref{fpratios}, the experimental data are also from the STAR~\cite{RHIC-star-mul-fluctuations-01} collaborations.
The data are for 0 - 5\% Au+Au collisions at various energies from $\sqrt{s_{NN}}=7.7$ to $200$ GeV, presented by open, half-filled blue stars.
Again, the SPheRIO results are given by open, half-filled red squares, for both average (left column) and event-by-event fluctuating (right column) ICs.
The HRG calculations are presented in purple dashed and dotted curves.}
\label{fpratios_sign}
\end{figure}

We carried out hydrodynamic simulations of Au+Au collisions based on the SPheRIO code for different centrality windows at different energies in accordance with the existing data of the BES program~\cite{LHC-na49-mul-fluctuations-01, RHIC-star-mul-fluctuations-01, RHIC-star-mul-fluctuations-02}.
The IC are generated by using NeXuS~\cite{nexus-1,nexus-rept}\footnote{This event generator has been updated and referred to as EPOS~\cite{epos-1,epos-2,epos-3}, but for the purpose of the present study, NeXuS is sufficient.}.
The results presented below are from simulations carried out for 977 events for 0 - 3.5\% Pb+Pb collisions at 8.8 GeV, as well as 756, 455, and 533 events for 0 - 5\% Au+Au collisions at 19.6, 62.4, and 200 GeV respectively.
For the sake of extracting the effects of event-by-event fluctuations, we also make use of the event-averaged IC, obtained by smoothing out the local density fluctuations for each centrality.

In Fig.~\ref{fpratios} and \ref{fpratios_sign}, we show the calculated dynamical fluctuations of particle ratios $p/\pi$, $K/\pi$, and $K/p$ at different energies. 
For instance, the quantity $\nu_{\rm dyn,p/\pi}$ measures the deviation in the ratios of $p/\pi$ with respect to those of an ideal statistical Poissonian distribution.
It is defined as,
\begin{eqnarray}
\nu_{\rm dyn,p/\pi} = \frac{\langle N_{p}(N_{p}-1)\rangle}{\langle N_{p}\rangle^{2}} + \frac{\langle N_{\pi}(N_{\pi}-1)\rangle}{\langle N_{\pi}\rangle^{2}} - 2\frac{\langle N_{p}N_{\pi}\rangle}{\langle N_{p}\rangle\langle N_{\pi}\rangle}\ .
\label{nudyn}
\end{eqnarray}
The results of hydrodynamic simulations by SPheRIO, and those of UrQMD as well as HRG models are presented together with the data from the NA49~\cite{LHC-na49-mul-fluctuations-01} and STAR~\cite{RHIC-star-mul-fluctuations-01} Collaborations.
In the case of SPheRIO, calculated results both with event-by-event fluctuating IC (in the right column denoted by ``w/ EbE") and event averaged IC (in the left column denoted by ``w/o EbE") are presented.
The error bars accompanying the hydrodynamical results correspond to the standard error related to the finite number of IC samples.

The SPheRIO results with event averaged IC show a quite reasonable agreement with those obtained by the static cases (HRG + resonance decays shown in continuous curves), and also with those from UrQMD\footnote{
When comparing against the particle list of UrQMD, SPheRIO considers all the baryons essentially up to 1.7 GeV and mesons up to 1 GeV.
Therefore, we believe that the difference does not quantitatively affect the discussions in the present study.}
This indicates that the corrections from the temporal expansion of the system are rather moderate.
However, it is interesting to note that the hydrodynamic effects for the $K/\pi$ cases appear slightly more significant regarding the others.
It is understood that the most dominant factor that leads to the above difference for the statistical model approach is the mass of specific particle species.
To be more specific, numerically, the contribution from the protons in Eq.~(\ref{nudyn}) is found to be less significant.
We also note that the resultant energy dependences and splitting among the isospin states of our model are more or less consistent with the experimental data, while the static HRG or UrQMD approaches give rather flat energy dependences.
On the other hand, for the $p/\pi$ case, SPheRIO results present systematic deviation in the lowest energy region, although the order of magnitude is still in accordance with the data.
We will come back to this point later.

When the event-by-event fluctuations are switched on, one finds that the calculated dynamical fluctuations are augmented.
For each term of Eq.~(\ref{nudyn}), both the numerator and denominator can be essentially cast into Eq.~(\ref{dNiNjeve}).
For the latter, the contributions due to the event-by-event fluctuations, on top of the thermal ones, are demonstrated in terms of covariance of thermal averages for different events.
As shown in Eq.~(\ref{dNiNjeve}), these covariances will be positive, as long as event-by-event multiplicity fluctuations of different species are positively correlated.
The overall effect, while one considers both the numerator and denominator, presented in various terms, gives rise to a slightly positive contribution.

Numerically, although the trend for lower energy Pb+Pb collisions is consistent with the data from NA49 Collaborations, the calculated dynamical fluctuations predominantly overestimate the experimental data.
Moreover, the obtained dynamical fluctuations are found to be significantly above the data and HRG model calculations.
The above difference is attributed to the event-by-event fluctuations in the IC generated by NeXuS.
To be more specific, it is speculated that the cause of the augmented dynamical fluctuations is the significant event-by-event local baryon density fluctuations associated with the baryon stopping presented primarily in low-energy events~\cite{Thakur:2016znw}.
This is manifested especially in the measurements shown in the top-right and bottom-right plots where protons are involved.
We understand that these fluctuations related to the baryon density are largely suppressed once one employs the event-average IC, and as a result, they are not observed in the case of the left column of Fig.~\ref{fpratios} and \ref{fpratios_sign}.
Furthermore, another possible cause of overwhelmed fluctuations might be related to the definition of centrality window. 
In fact, when the event-by-event fluctuations are switched on, an additional point, absent from the event averaged IC, comes into play.
To be specific, besides the baryon density fluctuations, significant multiplicity fluctuations may present even for a given impact parameter. 
However, we note that the experimental data seems to indicate that the STAR date on dynamical fluctuations of $p/\pi$ and $K/p$ are qualitatively different from those for $K/\pi$.
While the latter is mostly a monotonical function of energy, the former is characterized by a ``dip" at $\sqrt{s}\sim 20$ GeV.
This feature is not shown in the results of the HRG, UrQMD, and event-averaged hydrodynamical calculations.
It is somehow interesting to point out, in the case of event-by-event hydrodynamics, although not quantitatively, this tendency is reproduced owing to the elevated fluctuations presented in the low energy region.

For the present calculations, the definitions of centrality windows follow that of the impact parameters, while the experimentalists used multiplicity counts of charged tracks for given pseudo-rapidity region from the TPC detector.
It is understood that the use of impact parameters might potentially lead to more significant overall multiplicity fluctuations.
Therefore, to eliminate this potential ambiguity, we have carried out the calculations by using the definition of centrality window in terms of the overall multiplicity.
However, the resultant dynamical fluctuations of particle ratios are found almost identical in comparison with those presented in the right column of Fig.~\ref{fpratios} and \ref{fpratios_sign}.
Therefore we conclude that the overall multiplicity fluctuation does not play a significant role here for $\nu_{\rm dyn}$.
This probably can likely be attributed to the fact that, according to Eq.~(\ref{dNiNjeve}), the observable in question is normalized in terms of multiplicities for each species.

Also, we carry out calculations to show how the quantum ensemble considered in the present study is different from the scenario when one considers a classical ensemble.
The results are presented in Tab.~\ref{BoltzvsQuantum}.
There, the calculated dynamical fluctuations are further divided into different contributions, namely, those from thermal fluctuations and the rest associated with event-by-event initial fluctuations.
It is observed that the difference in thermal fluctuations between classical and quantum statistics is quite substantial.
The relative deviation is larger when light meson, such as $\pi$, is involved, which goes up and reaches 30\%.
Regarding the contributions from event-by-event fluctuating IC, on the other hand, the difference between classical and quantum statistics is not significant.
In the case where the magnitude of event-by-event fluctuations dominates, for instance, the $K/\pi$ fluctuations regarding the events at 200 GeV, the overall difference between the classical and quantum statistics is less significant.
This is because, for those cases, the event-by-event fluctuations play a crucial role in the overall contribution.
While on the other hand, when thermal fluctuations dominate, the overall difference due to classical or quantum statistics becomes more appreciable.

\begin{table}[htb]
\begin{center}
\scalebox{1.00}{\begin{tabular}{|c|c|c|c|c|c|c|c|c|c|c|}
\hline
\multirow{2}{*}{energy (GeV)} & \multirow{2}{*}{statistics}   & \multicolumn{3}{c|}{$\nu_{\rm dyn, K/ \pi}$ ($\times~10^{-4}$ )} & \multicolumn{3}{c|}{$\nu_{\rm dyn,p/ \pi}$  ($\times~10^{-4}$ )} & \multicolumn{3}{c|}{$\nu_{\rm dyn,K/p}$  ($\times~10^{-4}$ )}\\ \cline{3-11}  
&   & \multicolumn{1}{c|}{~EbE~} & \multicolumn{1}{c|}{thermal} & \multicolumn{1}{c|}{total} & \multicolumn{1}{c|}{~EbE~} & \multicolumn{1}{c|}{thermal} & \multicolumn{1}{c|}{total} & \multicolumn{1}{c|}{~EbE~} & \multicolumn{1}{c|}{thermal} & \multicolumn{1}{c|}{total} \\ \hline
\multirow{2}{*}{8.8} & \multicolumn{1}{c|}{$\rm{Boltzmann}$}  & $3.16 $    & $28.6  $  & $31.8   $ & $117 $ & $-8.88 $   & $108 $  & $128   $ & $23.7$  & $152$  \\ \cline{2-11} 
  & \multicolumn{1}{c|}{$\rm{quantum}$} & $3.04$ & $19.3 $  & $22.3 $ & $116$ & $ -10.7$  & $106 $ & $128$  & 
$22.4 $ & $150 $ \\ \hline
\multirow{2}{*}{19.6} 
 & \multicolumn{1}{c|}{$\rm{Boltzmann}$}  & $6.50 $    & $7.86 $  & $14.4   $ & $9.63 $ & $-4.06 $   & $5.57   $              & $21.6  $ & $7.27$  & $28.9$  \\ \cline{2-11}
  & \multicolumn{1}{c|}{$\rm{quantum}$} & $6.50$ & $5.74 $  & $12.2 $ & $9.72$ & $ -5.00$  & $4.72$ & $21.7$  & 
$6.37 $ & $28.1$ \\ \hline
\multirow{2}{*}{62.4} 
 & \multicolumn{1}{c|}{$\rm{Boltzmann}$}   & $5.97 $    & $5.91 $  & $11.9  $ & $17.6 $ & $-1.69 $   & $16.0  $              & $39.5    $ & $5.34$  & $44.9$  \\ \cline{2-11} 
  & \multicolumn{1}{c|}{$\rm{quantum}$} & $5.97$ & $4.34 $  & $10.3 $ & $17.7$ & $ -2.50$  & $15.2 $ & $39.5$  & 
$4.70 $ & $44.2 $ \\ \hline
\multirow{2}{*}{200} 
 & \multicolumn{1}{c|}{$\rm{Boltzmann}$}   & $10.5 $    & $6.03 $  & $16.5 $ & $15.6 $ & $-1.23 $   & $14.3  $ & $50.5  $ & $5.64 $  & $56.2 $  \\ \cline{2-11} 
  & \multicolumn{1}{c|}{$\rm{quantum}$} & $10.5 $ & $4.49 $  & $14.9 $ & $15.6 $ & $ -2.02 $  & $13.5 $ & $50.5$  & 
$4.96 $ & $55.5 $ \\ \hline
\end{tabular}}
\end{center}
\caption{The calculated dynamical fluctuations by considering Boltzmann as well as quantum statistics on the freeze-out surface.
The resulting contributions are divided into those of thermal as well as event-by-event (denoted by EbE) ones.}\label{BoltzvsQuantum}
\end{table}

\begin{figure}[htb!]
\begin{tabular}{ccc}
\begin{minipage}{170pt}
\centerline{\includegraphics[width=200pt]{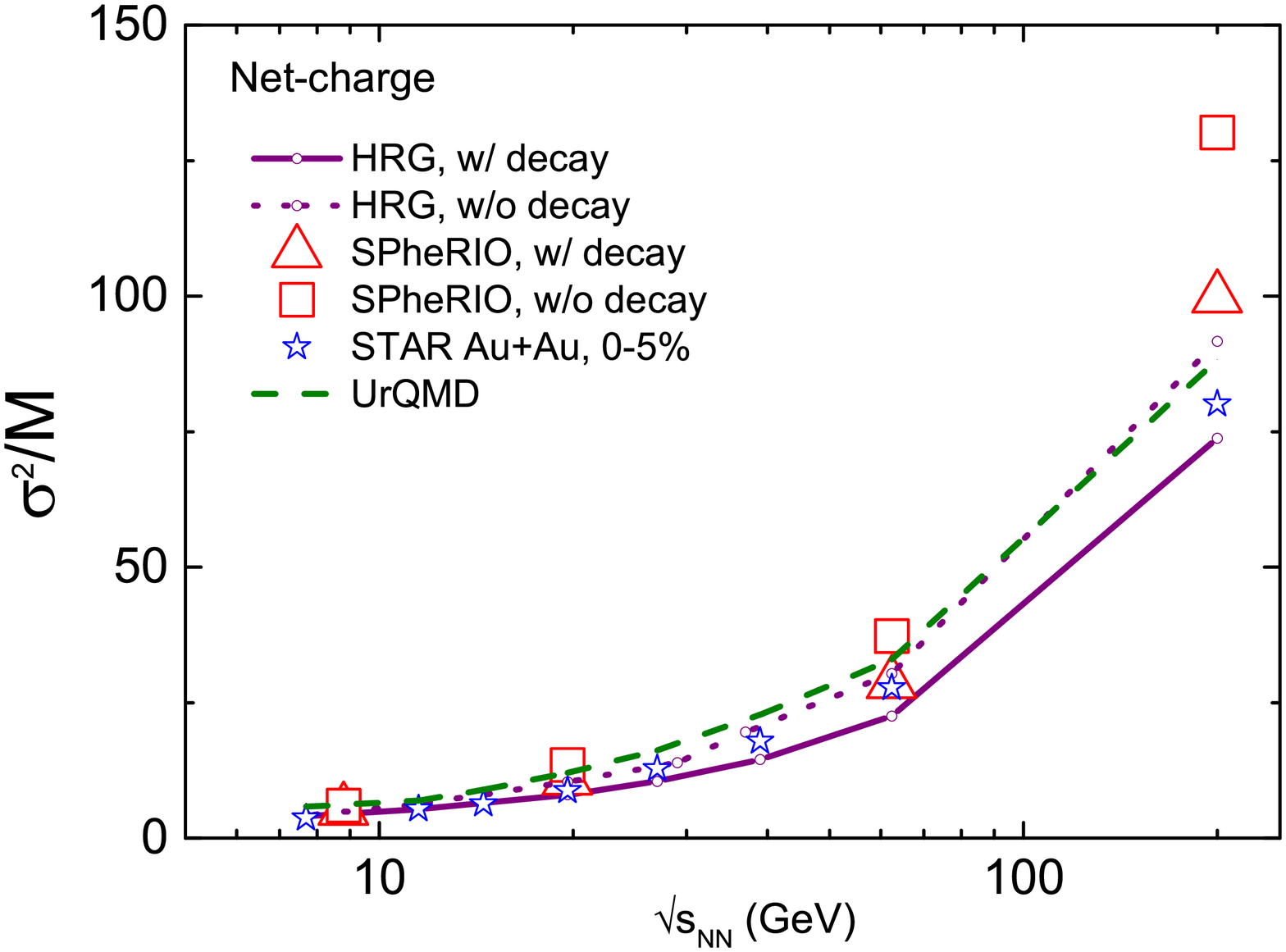}}
\end{minipage}
&
\begin{minipage}{170pt}
\centerline{\includegraphics[width=200pt]{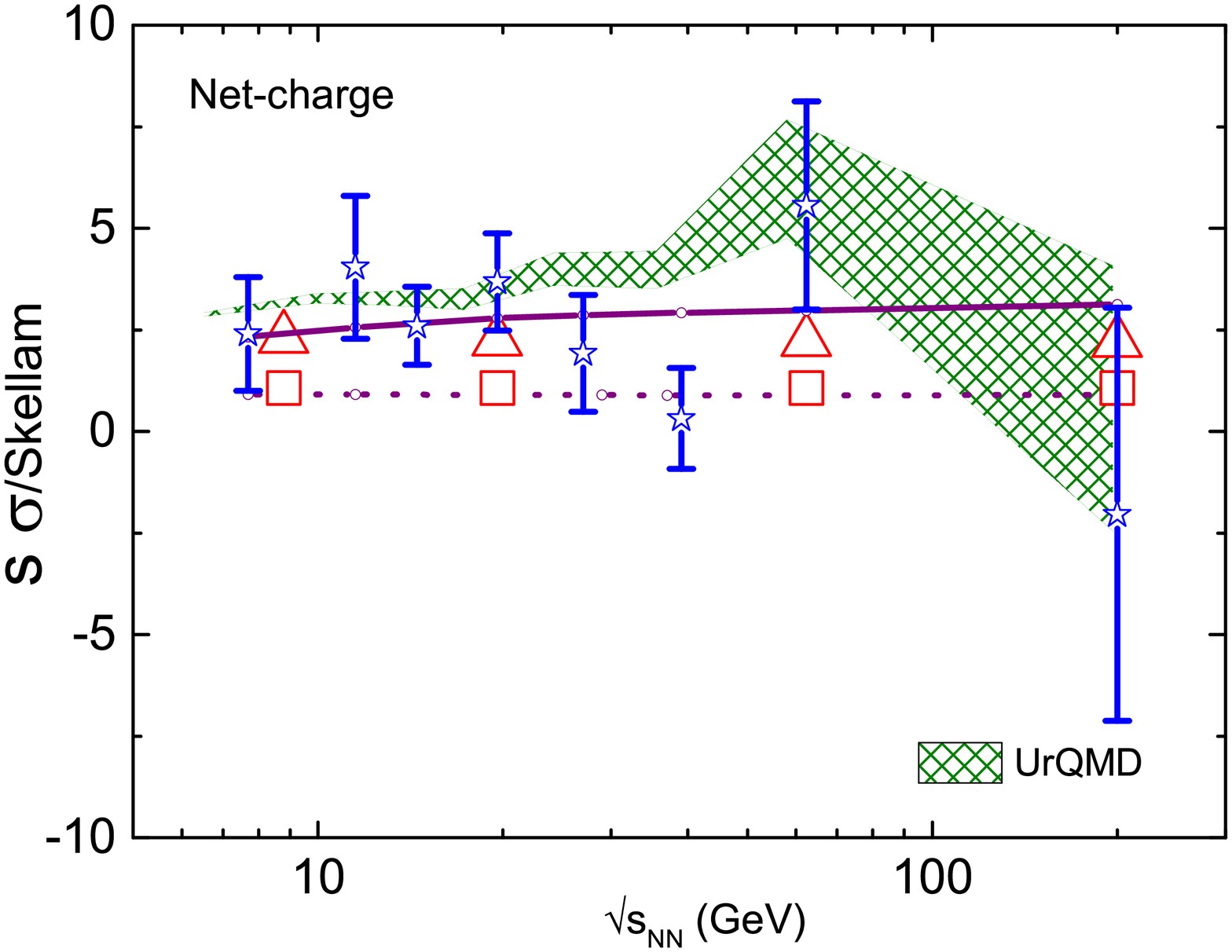}}
\end{minipage}
&
\begin{minipage}{170pt}
\centerline{\includegraphics[width=200pt]{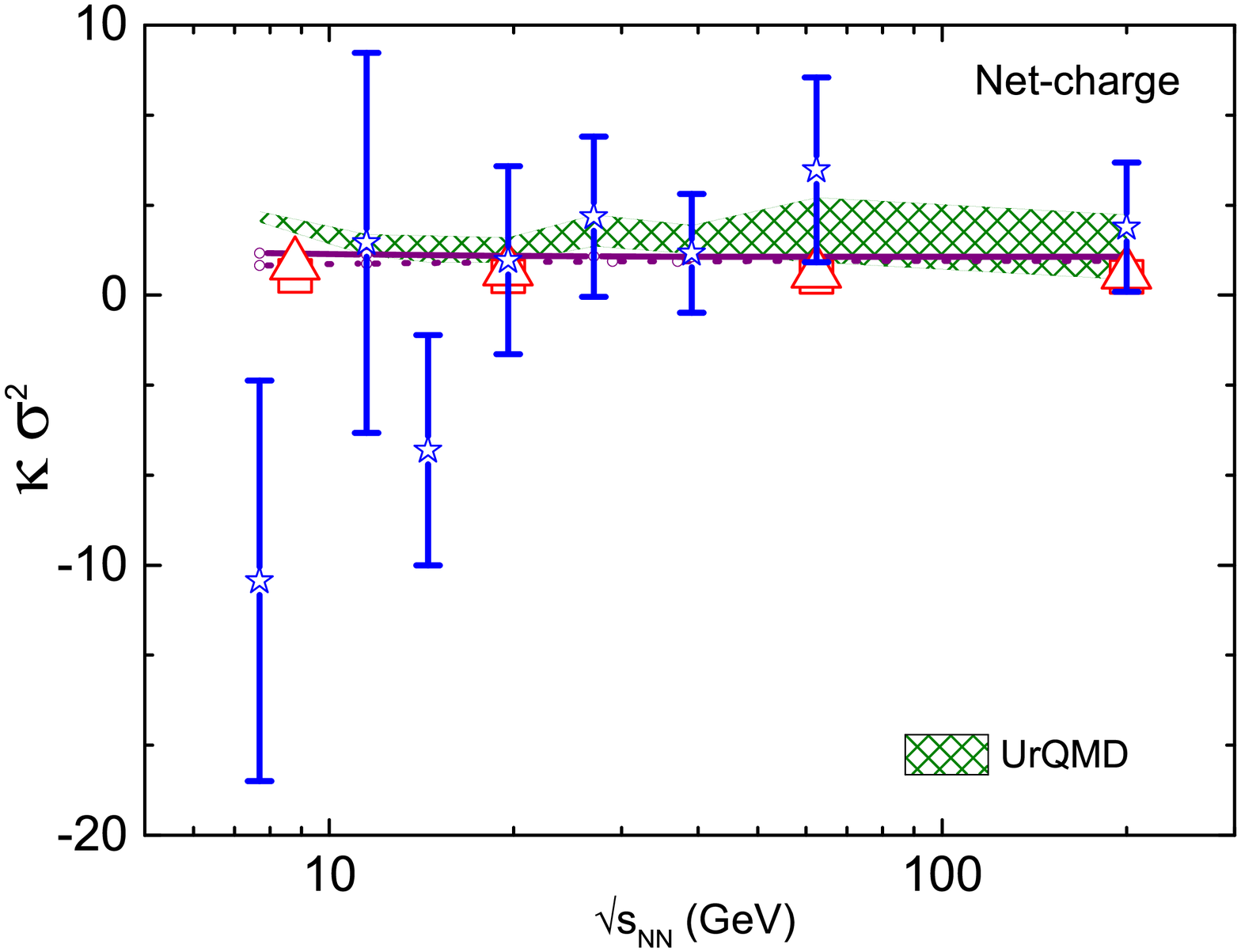}}
\end{minipage}
\\
\begin{minipage}{170pt}
\centerline{\includegraphics[width=200pt]{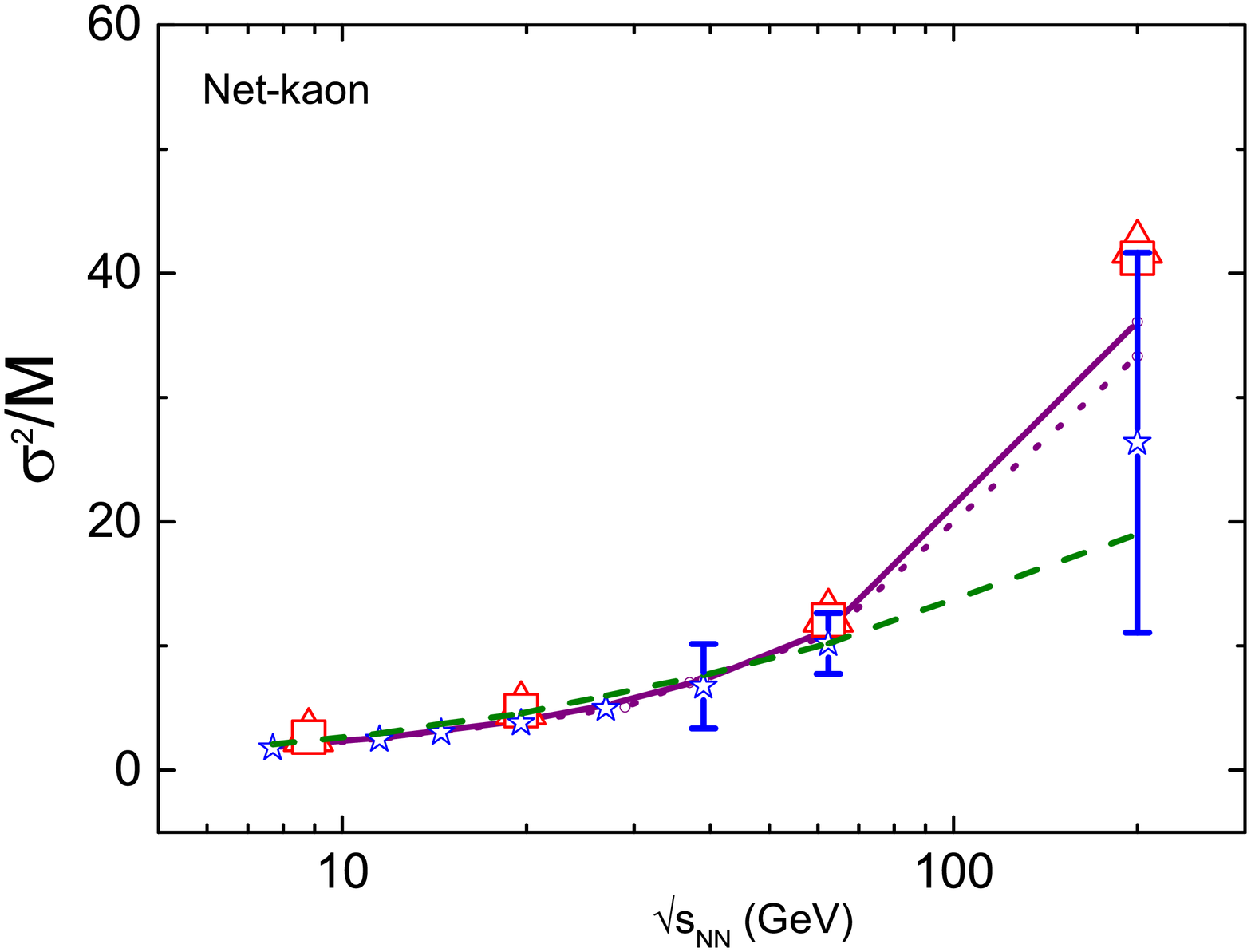}}
\end{minipage}
&
\begin{minipage}{170pt}
\centerline{\includegraphics[width=200pt]{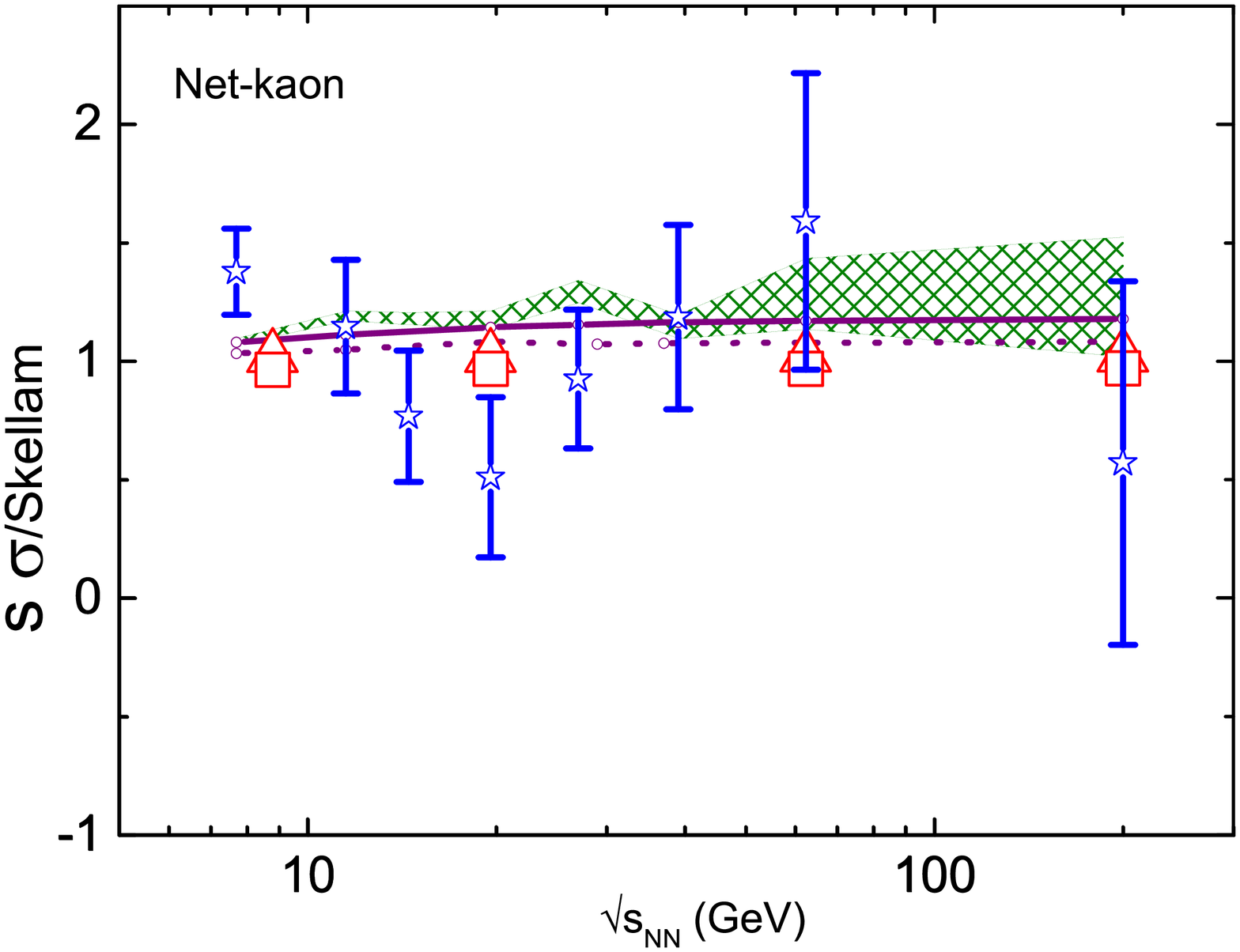}}
\end{minipage}
&
\begin{minipage}{170pt}
\centerline{\includegraphics[width=200pt]{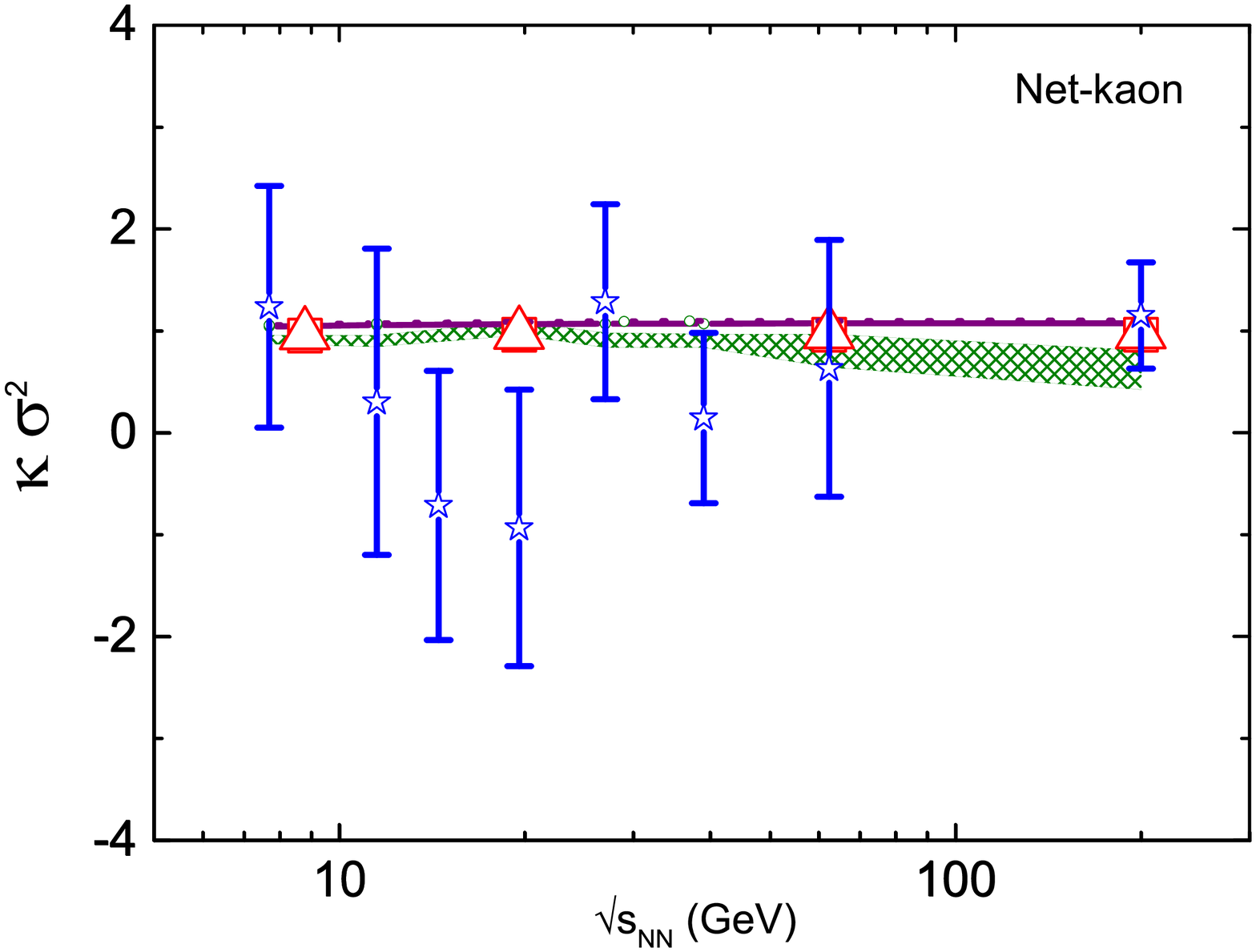}}
\end{minipage}
\\
\begin{minipage}{170pt}
\centerline{\includegraphics[width=200pt]{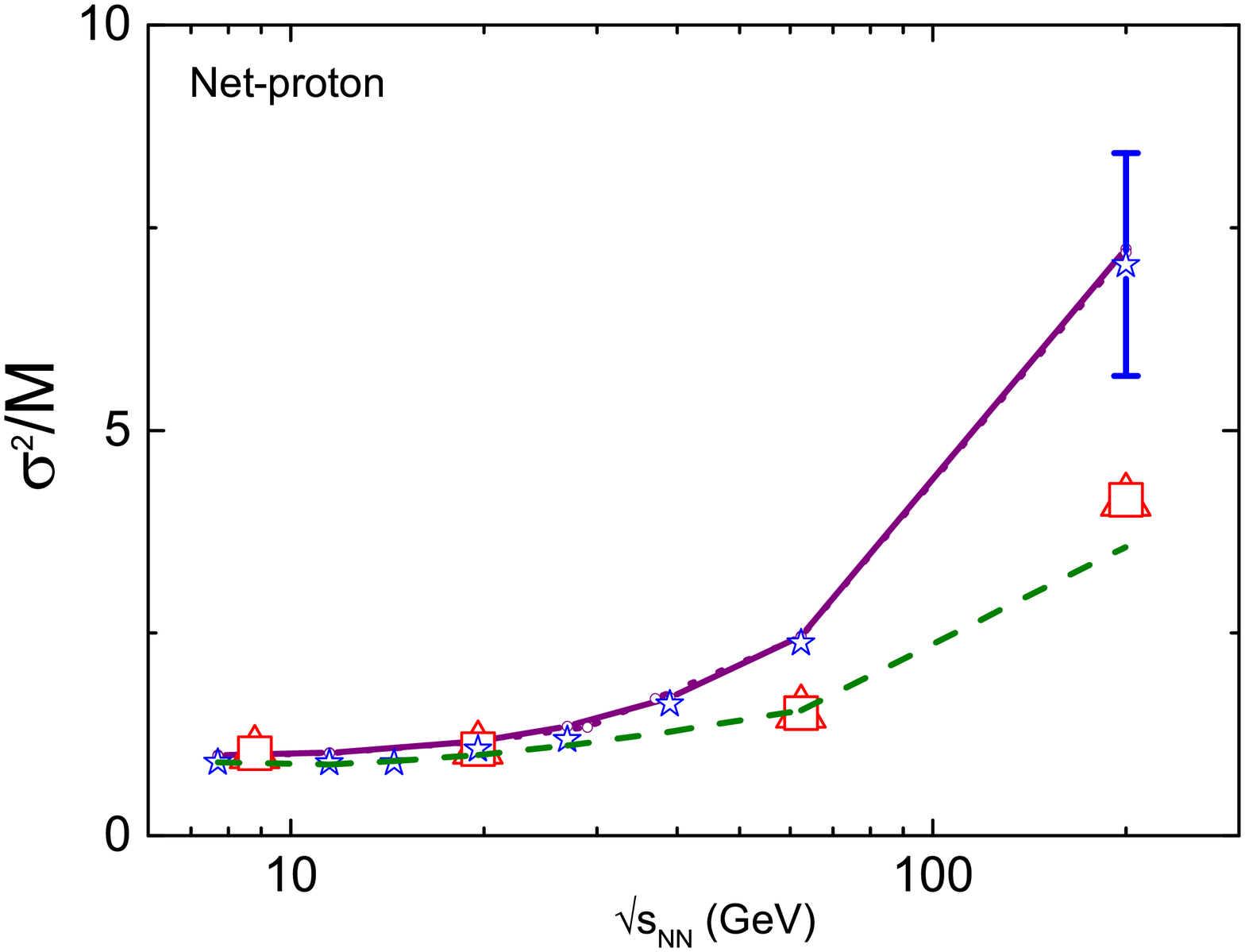}}
\end{minipage}
&
\begin{minipage}{170pt}
\centerline{\includegraphics[width=200pt]{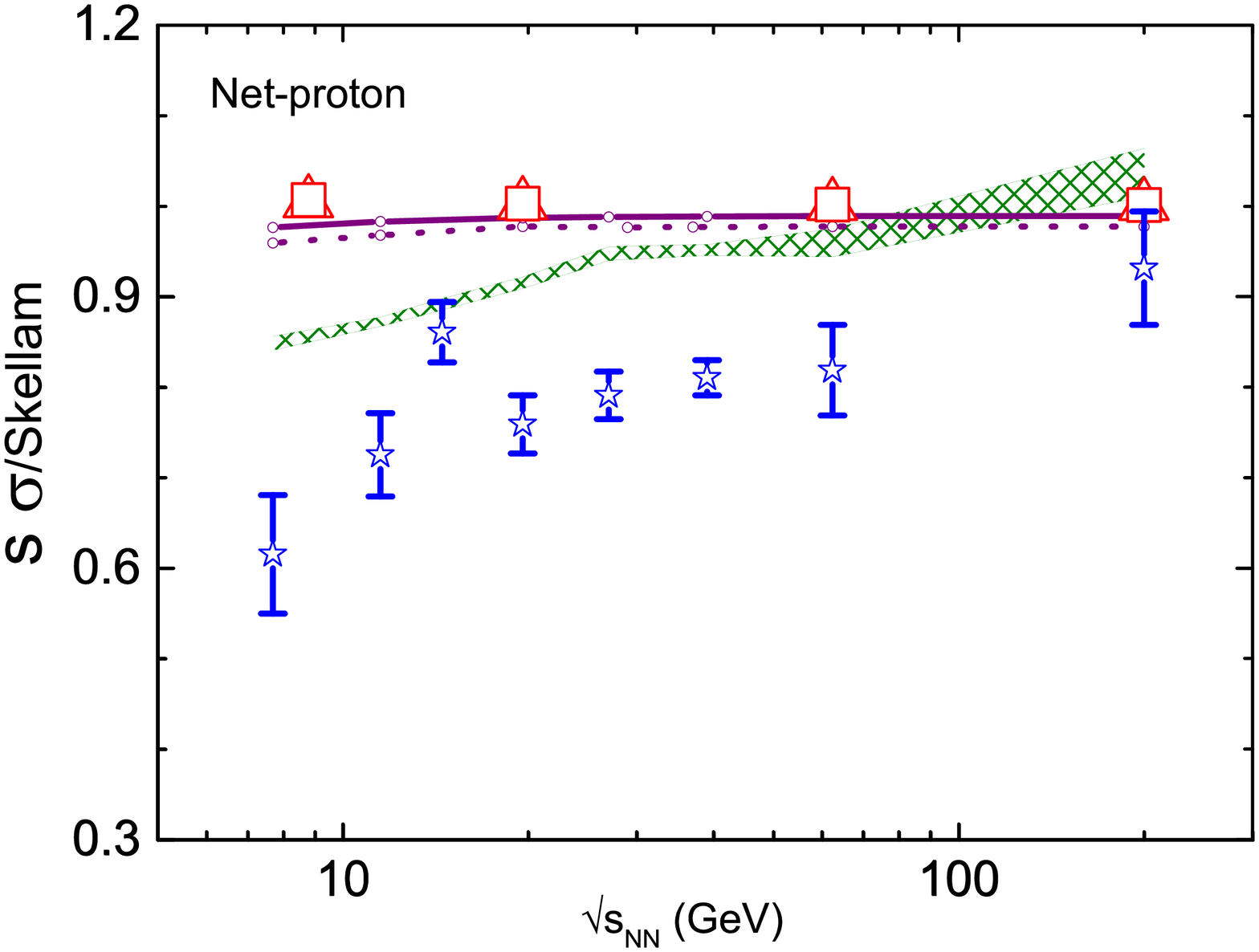}}
\end{minipage}
&
\begin{minipage}{170pt}
\centerline{\includegraphics[width=200pt]{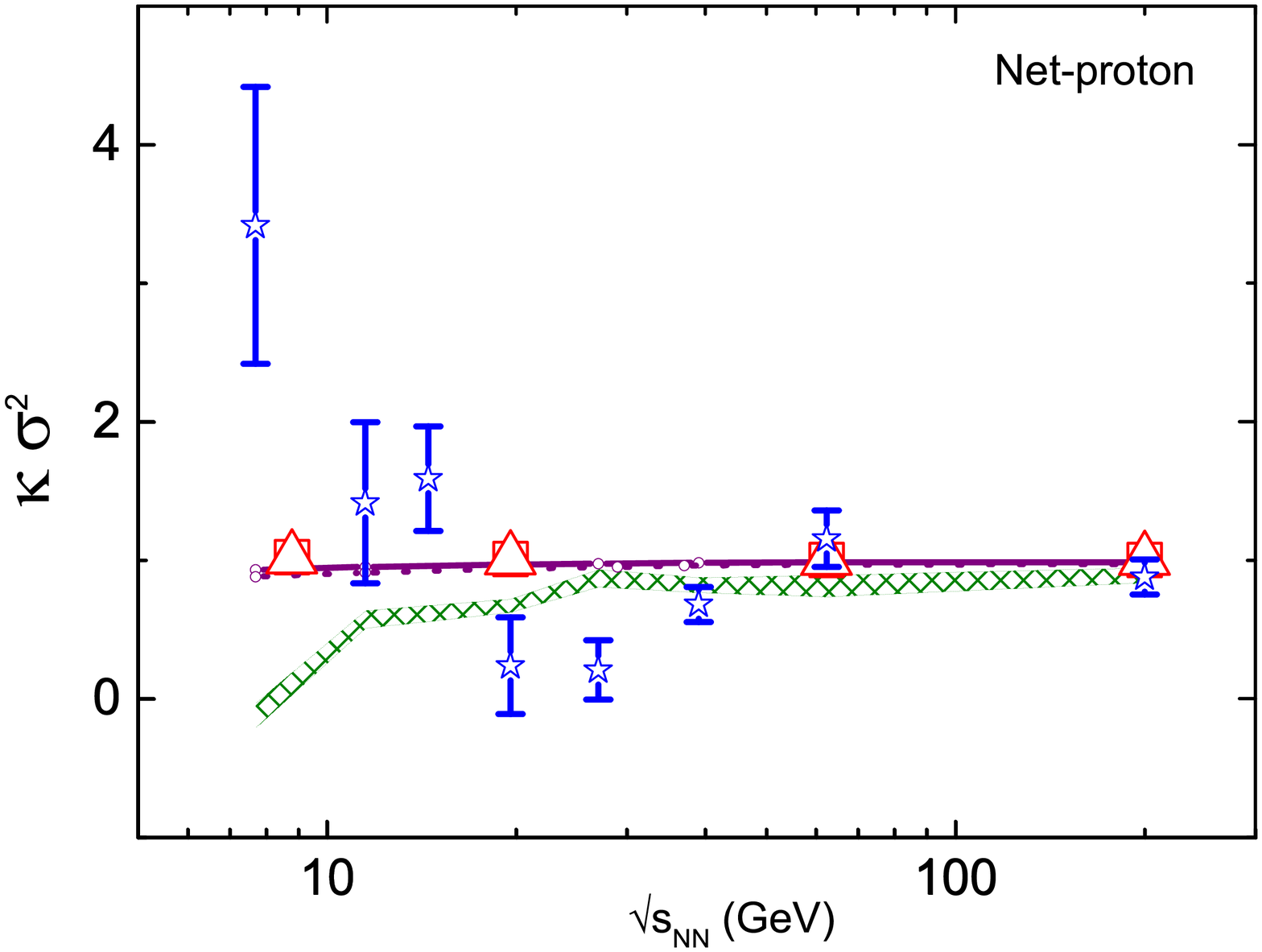}}
\end{minipage}
\end{tabular}
\caption{(Color online) The energy dependence of the higher moments of particle multiplicity.
The results for net-charge, net-kaon, and net-proton are shown in the top, middle, and bottom panels.
In the left, middle, and right columns, one presents the cumulant ratios $\sigma^2/M$, $S\sigma/\mathrm{Skellam}$, and $\kappa\sigma^2$, respectively. 
The STAR data~\cite{RHIC-star-mul-fluctuations-02} are for 0 - 5\% Au+Au collisions at various energies from $\sqrt{s_{NN}}=7.7$ to $200$ GeV, presented in open blue stars.
The corresponding SPheRIO results, with and without resonance decay, are shown in filled and open red circles, respectively.
Those obtained by UrQMD model calculations are displayed in dashed olive curves or cross-hatched area.
The HRG calculations are given in small open purple circles connected by solid curves.}
\label{fnetcharges}
\end{figure}

In Fig.~\ref{fnetcharges}, we present various cumulant ratios at different energies obtained by SPheRIO together with those by UrQMD and HRG models.
Here, the SPheRIO results are those of averaged ICs.
The STAR measurements~\cite{RHIC-star-mul-fluctuations-02} are for 0 - 5\% Au+Au collisions at various energies from $\sqrt{s_{NN}}=7.7$ to $200$ GeV.
As discussed above, the products $\kappa\sigma^2$ and $S\sigma$ are related to the ratios of particle number cumulants, which are identical to those of susceptibilities in a homogeneous system.
In particular, for an ideal Poissonian distribution, $S\sigma/\mathrm{Skellam}$ and $\kappa\sigma^2$ are both expected to be equal to 1.
For a hydrodynamic approach, the system is assumed to be in local equilibrium but not necessarily homogeneous. 
Numerically, the results from SPheRIO demonstrate a similar tendency as compared to those obtained by the HRG model.
These results are somewhat expected.
As mentioned before, for the smooth IC, the hydrodynamic calculations are not much different from the HRG ones since the freeze-out surface is relatively smooth, and its impact on particle fluctuations might be rather inconsequential.  
In the cases of $S\sigma/\mathrm{Skellam}$ and $\kappa\sigma^2$, unlike the UrQMD calculations, both HRG and hydrodynamical results indicate a less sensitive energy dependence.
For net-kaon fluctuations, both the HRG and hydrodynamical models give results consistent with the STAR measurements while considering the uncertainties. 
In comparison, for net-charge fluctuations, the observed energy dependence is reasonably captured by UrQMD simulations.
On the other hand, the measured $\kappa\sigma^2$ of net-proton decreases with decreasing energy, whereas none of the above models were able to reproduce such a trend.
As pointed out by the STAR Collaborations, non-monotonic behavior is observed in the energy dependence of the net-proton $\kappa\sigma^2$, subjected to further confirmation by improving the statistical and systematic uncertainties.
The presented results by hydrodynamical calculations based on GCE approach indicated that such non-monotonic feature does not come from the collective system expansion either thermal fluctuations.

\section{V. Further discussions and concluding remarks}

In this work, we studied some of the noncritical aspects of the multiplicity fluctuations in heavy-ion collisions by employing a hydrodynamic model. 
Apart from the critical behavior of the system near the critical point, there are many other sources which also contribute to the multiplicity fluctuations eventually observed experimentally.
In the HRG model, the effects of thermal fluctuations, finite volume correction, and resonance decay on the final multiplicities are taken into account.   
In this study, in addition to characteristics of the HRG approaches, we explore the fluctuations associated with the hydrodynamic freeze-out process. 
We further investigate how the present dynamical framework is affected by the IC by comparing the event-by-event generated ensembles to those resulting from a single smooth IC.  
It is also worth noting that we did not introduce any additional free parameter into the present hydrodynamic model, as the existing ones are determined in previous studies.
The obtained results are then compared to those of the HRG, UrQMD models, as well as the experimental data.
Overall, regarding the existing data, the results obtained by SPheRIO are reasonable in comparison with those by using different approaches.
In particular, it is observed that the event-by-event ICs may cause a sizable effect, especially at lower energies where the involved baryon density fluctuations might be significant. 
This, in turn, potentially implies a more stringent requirement for the event generator in terms of event-by-event fluctuations.
Moreover, it might be meaningful to carry out a more detailed analysis regarding a more realistic EoS focused on the region with finite baryon density. 
Furthermore, our results on the energy dependence of the cumulant ratios are mostly consistent with HRG and UrQMD model calculations.
Therefore, it is concluded that it is likely that experimentally observed non-monotonical behavior is not due to collective system expansion, either thermal fluctuations.

In our present study, we did not explicitly take into account the conserved charges.
We note that the magnitude of the fluctuations is reduced as one introduces more conservation laws~\cite{statistical-model-11}.
To be more rigorous, it is essential to explicitly include relevant conservation laws on an event-by-event basis when one studies the fluctuation.  
In particular, it has been shown that for a system of very few particle species, such a constraint in the evaluation of partition function is known to cause a remarkable suppression in particle number fluctuations~\cite{statistical-model-03}.
In the scenario of relativistic heavy-ion collisions, however, the total number of particle species is much larger, while on the other hand, there are a total of three conserved charges, namely, electric charge, baryon, and strangeness number.
As the number of multiplicities is much more significant to that of the conservation law, the effect of the latter might be less crucial.
As shown by full-fledged calculations carried out by using the HRG model~\cite{statistical-model-08}, the difference is less significant compared to the order of magnitude of the data.
Nevertheless, to properly implement the conservation of energy among other conserved charges is an essential aspect of the hydrodynamic model, which deserves attention.
As discussed above, the total energy discrepancy at the freeze-out surface becomes rather significant, especially for the case of event-by-event fluctuating ICs.
Besides, the employed hydrodynamical approach does not include the effect of viscosity.
Overall, we understand that the introduction of viscosity will further suppress the multiplicity fluctuations.
Also, viscosity is expected to have a significant impact on the collective flow of the high transverse momentum region.
Its effect on overall multiplicity fluctuations, however, may be less substantial in this regard.
Another relevant feature which is within the framework of hydrodynamics is the so-called continuous emission~\cite{sph-ce-01,sph-hbt-01}. 
In this context, since the hadronization takes place according to a given escape probability, the temperature at the freeze-out ``surface" is not a constant.
As a result, it gives rise to additional fluctuations in comparison to the case of the Cooper-Frye scenario.
Moreover, there are other hadronization scenarios employed in practice, where the freeze-out hypersurface is characterized by constant energy density, Knudsen number, etc.
It is interesting to investigate further how different freeze-out criterions affect the resultant multiplicity fluctuations.

In the literature, multiplicity fluctuations also have been investigated by using hydrodynamic approaches by other authors~\cite{hydro-fluctuations-02,hydro-fluctuations-03}.
In Ref.~\cite{hydro-fluctuations-03}, the cumulant ratios have been studied.
There, it was assumed that the multiplicity fluctuations during the hadron emission follow those of a GCE of a classical Maxwell-Boltzmann system, namely, the Poisson distribution.
Therefore, any resultant deviation from the latter is due to the effects of subsequential physical processes such as volume fluctuations, hadronic evolution, resonance decays, among others.
Our calculations have shown that the difference between classical and quantum ensemble can be substantial.
Other studies are focused on different aspects.
For instance, in Ref.~\cite{hydro-fluctuations-02}, the cause of the fluctuations is attributed to the quantum fluctuations in the vicinity of the critical point.
The latter is implemented by employing the spirit of the so-called $\sigma$ model where the fluctuations of a phenomenological $\sigma$ field were associated with those of emitted hadrons. 
The present study, on the other hand, is essentially based on the HRG model.
This is applied to every fluid element at the freeze-out surface, where the thermal fluctuations of a quantum GCE are accounted for, and the correlation functions are subsequently calculated analytically.

In this regard, a model which is aimed to probe relevant physics, meanwhile being able to reproduce the particle production with quantitatively correct numbers, shall be esteemed as more useful in the endeavor of BES program.
Although HRG models provide a seemingly reasonable description of the existing data, it is indeed meaningful to further incorporate the physics of critical phenomena explicitly into the present approach.
As discussed in the introduction, there are at least three relevant aspects.
First, the existence of a critical point may affect the EoS, even in the context of traditional hydrodynamics.
To study the impact on the multiplicity fluctuations regarding an EoS which carries explicit information on the critical point might be potentially interesting.
Secondly, a more fundamental approach involves the modification of the hydrodynamical equation of motion itself.
The chiral phase transition may directly impact the form of the hydrodynamical equation.
The above mentioned $\sigma$ model is an exciting possibility.
Besides, chiral hydro approaches implement the physics of the spontaneous symmetry breaking of a phenomenological chiral field in terms of the source term of the existing hydrodynamic equation.
The physics related to the critical slowing down may also affect the temporal evolution of the system on a fundamental level.
Last but not least, many other realistic factors should be implemented, especially when we intend to deal with experimental measurements. 
A further study in this direction is in progress.   

\section*{Acknowledgments}
We are thankful for valuable discussions with Nu Xu, Fr\'ed\'erique Grassi, and Matthew Luzum. 
We gratefully acknowledge the financial support from
Funda\c{c}\~ao de Amparo \`a Pesquisa do Estado de S\~ao Paulo (FAPESP),
Funda\c{c}\~ao de Amparo \`a Pesquisa do Estado do Rio de Janeiro (FAPERJ),
Conselho Nacional de Desenvolvimento Cient\'{\i}fico e Tecnol\'ogico (CNPq),
and Coordena\c{c}\~ao de Aperfei\c{c}oamento de Pessoal de N\'ivel Superior (CAPES).
A part of the work was developed under the project INCTFNA Proc. No. 464898/2014-5.
This research is also supported by the Center for Scientific Computing (NCC/GridUNESP) of the S\~ao Paulo State University (UNESP).

\clearpage

\section*{Appendix: Higher moments of the multiplicity distribution}

In this Appendix, we enumerate some of the expressions that are made use of in our numerical implementation.
Some of the formulae shown below have already be derived in the literature~\cite{statistical-model-03,statistical-model-04,statistical-model-08}, they are presented here for the sake of completeness.
 
By definition, the second, third and fourth order moments of multiplicity distribution can be written as
\begin{eqnarray}
\langle \Delta N_i  \Delta N_j \rangle&=& \langle N_i  N_j \rangle -\langle N_i \rangle \langle  N_j \rangle,
\\
\langle \Delta N_i  \Delta N_j \Delta N_k \rangle&=& \langle N_i  N_j N_K \rangle -\langle N_i  N_j \rangle \langle  N_k \rangle -\langle N_i  N_k \rangle \langle  N_j \rangle 
\nonumber\\
&-&\langle N_k  N_j \rangle \langle  N_i \rangle + 2 \langle N_i \rangle \langle  N_j \rangle \langle  N_k \rangle,
\\
\langle \Delta N_i  \Delta N_j \Delta N_k  \Delta N_l \rangle&=& \langle N_i  N_j N_K N_l\rangle -\langle N_i  N_j N_k \rangle  \langle  N_l \rangle -\langle N_i  N_k N_l \rangle \langle  N_j \rangle   \nonumber\\
&-&\langle N_i N_j  N_l \rangle \langle  N_k \rangle  - \langle N_j N_k  N_l \rangle \langle  N_i \rangle \nonumber\\
&+& \langle N_i  N_j \rangle \langle  N_l \rangle  \langle  N_k \rangle + \langle N_i  N_k  \rangle \langle  N_l \rangle \langle  N_j \rangle + \langle N_i  N_l \rangle \langle N_j \rangle \langle  N_k \rangle \nonumber\\
&+& \langle N_j  N_k \rangle \langle N_i \rangle \langle  N_l \rangle + \langle N_j  N_l \rangle \langle N_i \rangle \langle  N_k \rangle + \langle N_k  N_l \rangle \langle N_i \rangle \langle  N_j \rangle
\nonumber\\
&-&3 \langle N_i \rangle \langle  N_j \rangle \langle N_k \rangle \langle  N_l \rangle,
\end{eqnarray}
where the subscripts $i$, $j$, $k$, and $l$ represent the particle species. 

These quantities are closely associated with the higher order cumulants of particle number as follows~\cite{book-landau-5},
\begin{eqnarray}
\left\langle\left(\Delta N_{i}\right)^{3}\right\rangle &=& T^{2} \left(\frac{\partial^2 N_{i}}{\partial^2 \mu}\right)_{T},
\\
\left\langle\left(\Delta N_{i}\right)^{4}\right\rangle &-& 3 \left\langle\left(\Delta N_{i}\right)^{2}\right\rangle = T^{3} \left(\frac{\partial^3 N_{i}}{\partial^3 \mu}\right)_{T}.
\end{eqnarray}

By taking into considering that the covariance between different particle species vanishes, it is straightforward to find, with the aid of Eq.~(\ref{avgNi}),
\begin{eqnarray}
\left\langle \Delta N_{i} \Delta N_{j} \Delta N_{k}\right\rangle &=&\sum_{p} \left\langle \left(\Delta n_{p,i}\right)^{3} \right\rangle \nonumber\\
&=&\sum_{p}  2 \langle n_{p,i} \rangle  \left( 1+ \gamma_i \langle n_{p,i} \rangle \right)^2 - \langle n_{p,i} \rangle  \left( 1+ \gamma_i \langle n_{p,i} \rangle \right)\nonumber\\
&=& \sum_{p}   \langle n_{p,i} \rangle  \left( 1+ 3 \gamma_i \langle n_{p,i} \rangle + 2 \gamma_i^2  \langle n_{p,i} \rangle^2 \right),\\
\left\langle \Delta N_{i} \Delta  N_{j} \Delta  N_{k} \Delta  N_{l}\right\rangle 
&-& \left\langle\left(\Delta N_{i} \Delta N_{j} \right) \left(\Delta N_{k} \Delta N_{l} \right) \right\rangle 
- \left\langle\left(\Delta N_{i} \Delta N_{k} \right) \left(\Delta N_{j} \Delta N_{l} \right) \right\rangle
- \left\langle\left(\Delta N_{i} \Delta N_{l} \right) \left(\Delta N_{j} \Delta N_{k} \right) \right\rangle \nonumber\\
&=&\sum_{p} \left\langle \left(\Delta n_{p,i}\right)^{4} \right\rangle - 3 \left\langle \left(\Delta n_{p,i}\right)^{2} \right\rangle \nonumber\\
&=&\sum_{p} 6 \langle n_{p,i} \rangle  \left( 1+ \gamma_i \langle n_{p,i} \rangle \right)^3 - 6 \langle n_{p,i} \rangle  \left( 1+ \gamma_i \langle n_{p,i} \rangle \right)^2 + \langle n_{p,i} \rangle  \left( 1+ \gamma_i \langle n_{p,i} \rangle \right)\nonumber\\
&=& \sum_{p}  \langle n_{p,i} \rangle  \left( 1+ 7 \gamma_i \langle n_{p,i} \rangle + 12 \gamma_i^2 \langle n_{p,i} \rangle^2 + 6 \gamma_i^3 \langle n_{p,i} \rangle^3 \right).
\end{eqnarray}

In the RHIC BES data, skewness $S$ and kurtosis $\kappa$ are two quantities closely related to the measurements, and their definitions are closely related to the cumulants.
To be more specific, the following ratios are frequently being used
\begin{eqnarray}
\frac{\sigma^2}{M} &=& \frac{\left\langle (\Delta N)^2\right\rangle}{\left\langle N \right\rangle} ,\nonumber\\
S{\sigma} &=& \frac{\left\langle (\Delta N)^3\right\rangle}{\left\langle (\Delta N)^2 \right\rangle} ,\nonumber\\
\kappa{\sigma}^2 &=& \frac{\left\langle (\Delta N)^4\right\rangle - 3 \left\langle (\Delta N)^2\right\rangle^2}{\left\langle (\Delta N)^2 \right\rangle} .
\end{eqnarray}
The reason for the ratio combinations is that the above three quantities are identical to 1 in the case of ideal Poissonian distribution.

In practice, measurements are carried out for the net-particle multiplicity distribution regarding the above cumulant ratios.
For instance, for net-proton multiplicity distribution, one finds
\begin{eqnarray}
\frac{\sigma^2_{p-\bar p}}{M_{p-\bar p}} &=& \frac{\left\langle (\Delta N_{p-\bar p})^2\right\rangle}{\left\langle N_{p-\bar p} \right\rangle},\nonumber\\
S_{p-\bar p}{\sigma_{p-\bar p}} &=& \frac{\left\langle (\Delta N_{p-\bar p})^3\right\rangle}{\left\langle (\Delta N_{p-\bar p})^2 \right\rangle},\nonumber\\
\kappa_{p-\bar p}{\sigma^2_{p-\bar p}} &=& \frac{\left\langle (\Delta N_{p-\bar p})^4\right\rangle - 3 \left\langle (\Delta N_{p-\bar p})^2\right\rangle^2}{\left\langle (\Delta N_{p-\bar p})^2 \right\rangle} ,
\end{eqnarray}
where
\begin{eqnarray}
\left\langle N_{p-\bar p} \right\rangle=&&\langle N_{p} \rangle - \langle N_{\bar{p}} \rangle,
\\
\langle (\Delta N_{p-\bar{p}})^2 \rangle=&& \langle (\Delta N_{p})^2 \rangle +  \langle (\Delta N_{\bar{p}})^2 \rangle -  2 \langle \Delta N_{p} \Delta N_{\bar{p}}\rangle,
\\
\langle (\Delta N_{p-\bar{p}})^3 \rangle= &&
 \langle (\Delta N_{p} )^3 \rangle - \langle (\Delta N_{\bar{p}} )^3 \rangle -
3 \langle (\Delta N_{p})^2 \Delta N_{\bar{p}} \rangle + 3 \langle \Delta N_{p} (\Delta N_{\bar{p}})^2 \rangle,
\\
\langle (\Delta N_{p-\bar{p}})^4 \rangle -&& 3 \langle (\Delta N_{p-\bar{p}})^2 \rangle= \langle (\Delta N_{p})^4 \rangle - 3 \langle (\Delta N_{p})^2 \rangle^2
+\langle (\Delta N_{\bar{p}})^4 \rangle - 3 \langle (\Delta N_{\bar{p}})^2 \rangle^2
\nonumber\\
-&&4 (\langle (\Delta N_{p})^3 \Delta N_{\bar{p}}\rangle - 3 \langle (\Delta N_{p})^2 \rangle \langle \Delta N_{p} \Delta N_{\bar{p}}\rangle)
\nonumber\\
+&&6 (\langle (\Delta N_{p})^2 (\Delta N_{\bar{p}})^2 \rangle-2 \langle \Delta N_{p} \Delta N_{\bar{p}}\rangle^2 -  \langle (\Delta N_{p})^2 \rangle \langle (\Delta N_{\bar{p}})^2\rangle)
\nonumber\\
-&&4 (\langle (\Delta N_{\bar{p}})^3 \Delta N_{p}\rangle - 3 \langle (\Delta N_{\bar{p}})^2 \rangle \langle \Delta N_{p} \Delta N_{\bar{p}}\rangle).
\end{eqnarray}

In the case of ideal Poissonian distribution, it is straightforward to show that
\begin{eqnarray}
\frac{\sigma^2_{p-\bar p}}{M_{p-\bar p}} &\to&  \frac{\left\langle N_p\right\rangle + \left\langle N_{\bar p}\right\rangle}{\left\langle N_p\right\rangle - \left\langle N_{\bar p}\right\rangle} ,\nonumber\\
S_{p-\bar p}{\sigma_{p-\bar p}} &\to& \frac{\left\langle N_p\right\rangle - \left\langle N_{\bar p}\right\rangle}{\left\langle N_p\right\rangle + \left\langle N_{\bar p}\right\rangle} ,\nonumber\\
\kappa_{p-\bar p}{\sigma^2_{p-\bar p}} &\to& 1  .
\end{eqnarray}
while the emissions of protons and anti-protons are treated as independent.

As one further considers event-by-event fluctuating ICs, the above quantities are further modified to include the fluctuations between different events. 
For a total of $n$ events, $E \equiv \{E_1, E_2, \cdots, E_n \} $, one has
\begin{eqnarray}
\langle (\Delta N_i) (\Delta N_j)\rangle_{E} =&&\frac{1}{n}  \left[  \langle (\Delta N_i) (\Delta N_j)\rangle_{E_1} +  \langle (\Delta N_i) (\Delta N_j)\rangle_{E_2} + \cdots + \langle (\Delta N_i) (\Delta N_j)\rangle_{E_n} \right] \nonumber\\
+&& \frac{1}{n}  \left[ (\langle N_i \rangle_{E_1}- \langle N_i \rangle_{E} ) (\langle N_j \rangle_{E_1}- \langle N_j \rangle_{E} ) +  (\langle N_i \rangle_{E_2}- \langle N_i \rangle_{E} ) (\langle N_j \rangle_{E_2}- \langle N_j \rangle_{E} ) \right. \nonumber\\
&&+\cdots + \left. (\langle N_i \rangle_{E_n}- \langle N_i \rangle_{E} ) (\langle N_j \rangle_{E_n}- \langle N_j \rangle_{E} ) \right] . \label{dNiNjeve}
\end{eqnarray}
Here $\langle \cdots \rangle_{E_k}$ indicates the ensemble average discussed above, therefore $\Delta N_i$ in the first term on the r.h.s. of the above expression is evaluated with respect to the ensemble average for a given event $k$.
However, $\langle \cdots \rangle_E$ stands for the event average, in the sense that $\Delta N_i$ on the l.h.s. is regarding the event average of ensemble ones.

In terms of SPH degree of freedom, the above results can be rewritten as follows
\begin{eqnarray}
\langle \Delta N_{i}\Delta N_{j} \Delta N_{k} \rangle &=& \int p_{\bot}dp_{\bot}dy d\phi \sum_j \frac{\nu_j n_{j\mu}p^{\mu}}{s_j|n_{j\rho}u_j^{\rho}|}\theta(u_{j\delta}p^{\delta}) v_{i}^{3}(u_{j\nu}p^{\nu}, x),\\\label{Ncum3SPH}
\langle \Delta N_{i}\Delta N_{j} \Delta N_{k} \Delta N_{l} \rangle &=& \int p_{\bot}dp_{\bot}dy d\phi \sum_j \frac{\nu_j n_{j\mu}p^{\mu}}{s_j|n_{j\rho}u_j^{\rho}|}\theta(u_{j\delta}p^{\delta})  v_{i}^{4}(u_{j\nu}p^{\nu}, x),\label{Ncum4SPH}
\end{eqnarray}
where
\begin{eqnarray}
 v_{i}^{3}(u_{j\nu}p^{\nu}, x)&=& n_i(u_{j\nu}p^{\nu}, x)  \left( 1+ 3 \gamma_i n_{i}(u_{j\nu}p^{\nu}, x)  + 2 \gamma_i^2  n_{i}^{2}(u_{j\nu}p^{\nu}, x) \right),  \nonumber\\
 v_{i}^{4}(u_{j\nu}p^{\nu}, x)&=& n_{i}(u_{j\nu}p^{\nu}, x)  \left( 1+ 7 \gamma_i n_{i}(u_{j\nu}p^{\nu}, x) + 12 \gamma_i^2  n_{i}^{2}(u_{j\nu}p^{\nu}, x) + 6 \gamma_i^3 n_{i}^{3}(u_{j\nu}p^{\nu}, x)  \right) .
\end{eqnarray}

Now, when resonance decay is considered, the three- and four-particle correlators entirely due to resonance decay can be evaluated by making use of the generating function defined in Eq.~(\ref{resonanceDecayGenerator}) 
\begin{eqnarray}
\overline{N_i N_j N_k} &\equiv& \sum_R\langle N_i N_j N_k\rangle_R+ \sum_{R \ne R'}  \langle N_i N_j N_k \rangle_{R,R'} + \sum_{R \ne R'\ne R''} \langle N_i N_j N_k\rangle_{R,R',R''}  =\lambda_i\frac{\partial }{\partial \lambda_i}\left(\lambda_j\frac{\partial }{\partial \lambda_j} \left(\lambda_k\frac{\partial }{\partial \lambda_k} G \right) \right)  \nonumber\\
&=&\sum_{R}  \left[ N_{R}  \langle n_i n_j n_k \rangle_{R} + N_{R} (N_{R} -1) \left( \langle n_i n_j \rangle_{R} \langle n_k \rangle_{R}+ \langle n_i n_k \rangle_{R} \langle n_j \rangle_{R} + \langle n_k n_j \rangle_{R} \langle n_i \rangle_{R} \right) \right. \nonumber\\
&+& \left. N_{R} (N_{R} -1) (N_{R} -2) \langle n_i \rangle_{R} \langle n_j \rangle_{R} \langle n_k \rangle_{R} \right]  \nonumber\\
&+& \sum_{R \ne R'} N_{R'} \left( N_{R}  \langle n_i n_j \rangle_{R} + N_{R} (N_{R} -1) \langle n_i \rangle_{R} \langle n_j \rangle_{R}  \right) \langle n_k \rangle_{R'} \nonumber\\ 
&+& \sum_{R \ne R'} N_{R'} \left( N_{R} \langle n_i n_k \rangle_{R} + N_{R} (N_{R} -1) \langle n_i \rangle_{R} \langle n_k \rangle_{R}  \right) \langle n_j \rangle_{R'} \nonumber\\
&+& \sum_{R \ne R'} N_{R'}\left( N_{R}  \langle n_j n_k \rangle_{R} + N_{R} (N_{R} -1) \langle n_j \rangle_{R} \langle n_k \rangle_{R}  \right) \langle n_i \rangle_{R'} \nonumber\\
&+& \sum_{R \ne R' \ne R''} N_{R} N_{R'} N_{R''} \langle n_i \rangle_{R} \langle n_j \rangle_{R'}  \langle n_k \rangle_{R''}. \label{res3}
\end{eqnarray}

\begin{eqnarray}
\overline{N_i N_j N_k N_l} &\equiv& \sum_R\langle N_i N_j N_k N_l\rangle_R+ \sum_{R \ne R'}  \langle N_i N_j N_k N_l \rangle_{R,R'} + \sum_{R \ne R'\ne R''} \langle N_i N_j N_k N_l \rangle_{R,R',R''} \nonumber\\
&+& \sum_{R \ne R'\ne R''\ne R'''} \langle N_i N_j N_k N_l \rangle_{R,R',R'',R'''}= \lambda_i\frac{\partial }{\partial \lambda_i}\left(\lambda_j\frac{\partial }{\partial \lambda_j} \left(\lambda_k\frac{\partial }{\partial \lambda_k} \left(\lambda_l\frac{\partial }{\partial \lambda_l} G \right) \right) \right)  \nonumber\\
&=& \sum_{R} \left[ N_R \langle n_i n_j n_k n_l\rangle_{R} + N_{R} (N_{R} -1) \left( \langle n_i n_j n_k\rangle_{R}  \langle n_l\rangle_{R} \right) \right. \nonumber\\
&+& N_{R} (N_{R} -1) \left( \langle n_i n_j n_l\rangle_{R}  \langle n_k\rangle_{R} + \langle n_i n_j \rangle_{R}  \langle n_l n_k\rangle_{R} \right) \nonumber\\
&+& N_{R} (N_{R} -1) \left( \langle n_i n_k n_l\rangle_{R}  \langle n_j\rangle_{R} + \langle n_i n_k \rangle_{R}  \langle n_l n_j\rangle_{R}  \right) \nonumber\\
&+& N_{R} (N_{R} -1) \left( \langle n_j n_k n_l\rangle_{R}  \langle n_i\rangle_{R} + \langle n_j n_k \rangle_{R}  \langle n_l n_i \rangle_{R}  \right) \nonumber\\
&+& N_{R} (N_{R} -1) (N_{R} -2) \left( \langle n_i n_j \rangle_{R}  \langle n_k\rangle_{R} \langle n_l\rangle_{R} + \langle n_i n_k \rangle_{R}  \langle n_j\rangle_{R} \langle n_l\rangle_{R} + \langle n_j n_k \rangle_{R}  \langle n_i\rangle_{R} \langle n_l\rangle_{R} \right) \nonumber\\
&+& N_{R} (N_{R} -1) (N_{R} -2) \left( \langle n_i n_l \rangle_{R}  \langle n_j\rangle_{R} \langle n_k\rangle_{R} + \langle n_j n_l \rangle_{R}  \langle n_i\rangle_{R} \langle n_k\rangle_{R} + \langle n_k n_l \rangle_{R}  \langle n_i\rangle_{R} \langle n_j\rangle_{R} \right)\nonumber\\
&+& \left. N_{R} (N_{R} -1) (N_{R} -2) (N_{R} -3) \langle n_i \rangle_{R} \langle n_j \rangle_{R} \langle n_k \rangle_{R} \langle n_l \rangle_{R} \right] \nonumber\\
&+& \sum_{R \ne R'} N_{R'} \left[ N_{R}  \langle n_i n_j n_k \rangle_{R} + N_{R} (N_{R} -1) \left( \langle n_i n_j \rangle_{R} \langle n_k \rangle_{R}+ \langle n_i n_k \rangle_{R} \langle n_j \rangle_{R} + \langle n_j n_k \rangle_{R} \langle n_i \rangle_{R}\right)  \right. \nonumber\\ 
&+& \left. N_{R} (N_{R} -1) (N_{R} -2) \langle n_i \rangle_{R} \langle n_j \rangle_{R} \langle n_k \rangle_{R} \right] \langle n_l \rangle_{R'} \nonumber\\
&+& \sum_{R \ne R'} N_{R'} \left[ N_{R}  \langle n_i n_j n_l \rangle_{R} + N_{R} (N_{R} -1) \left( \langle n_i n_j \rangle_{R} \langle n_l \rangle_{R}+ \langle n_i n_l \rangle_{R} \langle n_j \rangle_{R} + \langle n_j n_l \rangle_{R} \langle n_i \rangle_{R}\right)  \right. \nonumber\\ 
&+& \left. N_{R} (N_{R} -1) (N_{R} -2) \langle n_i \rangle_{R} \langle n_j \rangle_{R} \langle n_l \rangle_{R} \right] \langle n_k \rangle_{R'} \nonumber\\
&+& \sum_{R \ne R'} N_{R'} \left[ N_{R}  \langle n_i n_k n_l \rangle_{R} + N_{R} (N_{R} -1) \left( \langle n_i n_k \rangle_{R} \langle n_l \rangle_{R}+ \langle n_i n_l \rangle_{R} \langle n_k \rangle_{R} + \langle n_k n_l \rangle_{R} \langle n_i \rangle_{R}\right)  \right. \nonumber\\ 
&+& \left. N_{R} (N_{R} -1) (N_{R} -2) \langle n_i \rangle_{R} \langle n_k \rangle_{R} \langle n_l \rangle_{R} \right] \langle n_j \rangle_{R'} \nonumber\\
&+& \sum_{R \ne R'} N_{R'} \left[ N_{R}  \langle n_j n_k n_l \rangle_{R} + N_{R} (N_{R} -1) \left( \langle n_j n_k \rangle_{R} \langle n_l \rangle_{R}+ \langle n_j n_l \rangle_{R} \langle n_k \rangle_{R} + \langle n_k n_l \rangle_{R} \langle n_j \rangle_{R}\right)  \right. \nonumber\\ 
&+& \left. N_{R} (N_{R} -1) (N_{R} -2) \langle n_j \rangle_{R} \langle n_k \rangle_{R} \langle n_l \rangle_{R} \right] \langle n_i \rangle_{R'} \nonumber\\
&+& \sum_{R \ne R'}  \left[ N_{R} \langle n_i n_j \rangle_{R} + N_{R} (N_{R} -1) \langle n_i \rangle_{R} \langle n_j \rangle_{R}  \right] \left[ N_{R'} \langle n_k n_l \rangle_{R'} + N_{R'} (N_{R'} -1) \langle n_k \rangle_{R'} \langle n_l \rangle_{R'}  \right] \nonumber\\
&+& \sum_{R \ne R'}  \left[ N_{R} \langle n_i n_k \rangle_{R} + N_{R} (N_{R} -1) \langle n_i \rangle_{R} \langle n_k \rangle_{R}  \right] \left[ N_{R'} \langle n_j n_l \rangle_{R'} + N_{R'} (N_{R'} -1) \langle n_j \rangle_{R'} \langle n_l \rangle_{R'}  \right] \nonumber\\
&+& \sum_{R \ne R'}  \left[ N_{R} \langle n_j n_k \rangle_{R} + N_{R} (N_{R} -1) \langle n_j \rangle_{R} \langle n_k \rangle_{R}  \right] \left[ N_{R'} \langle n_i n_l \rangle_{R'} + N_{R'} (N_{R'} -1) \langle n_i \rangle_{R'} \langle n_l \rangle_{R'}  \right] \nonumber\\
&+& \sum_{R \ne R' \ne R''} N_{R'} N_{R''}  \left[N_{R} \langle n_i n_j \rangle_{R} + N_{R} (N_{R} -1) \langle n_i \rangle_{R} \langle n_j \rangle_{R}  \right] \langle n_k \rangle_{R'} \langle n_l \rangle_{R''} \nonumber\\
&+& \sum_{R \ne R' \ne R''} N_{R'} N_{R''}  \left[N_{R} \langle n_i n_k \rangle_{R} + N_{R} (N_{R} -1) \langle n_i \rangle_{R} \langle n_k \rangle_{R}  \right] \langle n_j \rangle_{R'} \langle n_l \rangle_{R''} \nonumber\\
&+& \sum_{R \ne R' \ne R''} N_{R'} N_{R''}  \left[N_{R} \langle n_i n_l \rangle_{R} + N_{R} (N_{R} -1) \langle n_i \rangle_{R} \langle n_l \rangle_{R}  \right] \langle n_j \rangle_{R'} \langle n_k \rangle_{R''} \nonumber\\
&+& \sum_{R \ne R' \ne R''} N_{R'} N_{R''}  \left[N_{R} \langle n_j n_k \rangle_{R} + N_{R} (N_{R} -1) \langle n_j \rangle_{R} \langle n_k \rangle_{R}  \right] \langle n_i \rangle_{R'} \langle n_l \rangle_{R''} \nonumber\\
&+& \sum_{R \ne R' \ne R''} N_{R'} N_{R''}  \left[N_{R} \langle n_j n_l \rangle_{R} + N_{R} (N_{R} -1) \langle n_j \rangle_{R} \langle n_l \rangle_{R}  \right] \langle n_i \rangle_{R'} \langle n_k \rangle_{R''} \nonumber\\
&+& \sum_{R \ne R' \ne R''} N_{R'} N_{R''}  \left[N_{R} \langle n_k n_l \rangle_{R} + N_{R} (N_{R} -1) \langle n_k \rangle_{R} \langle n_l \rangle_{R}  \right] \langle n_i \rangle_{R'} \langle n_j \rangle_{R''} \nonumber\\
&+& \sum_{R \ne R' \ne R'' \ne R'''} N_{R} N_{R'} N_{R''} N_{R'''} \langle n_i \rangle_{R} \langle n_j \rangle_{R'}  \langle n_k \rangle_{R''} \langle n_l \rangle_{R'''}.   \label{res4}
\end{eqnarray}

Now by taking into consideration the primordial particles created before the resonance decay, namely,
\begin{eqnarray}
\langle \Delta N_i  \rangle = \langle \Delta N_{i}^{*} \rangle + \sum_{R}  \langle N_{R} \rangle \sum_r b_r^Rn_{i,r}^R \equiv \langle \Delta N_{i}^{*} \rangle + \sum_{R}  \langle N_{R} \rangle \langle n_i \rangle_{R} ,
\end{eqnarray}
where the terms with the superscript ``$*$" indicate the corresponding primordial quantities before the decay process.
Subsequently, the covariance between the particles of species $i$ and $j$ after the resonance decay is
\begin{eqnarray}
\langle \Delta N_i  \Delta N_j \rangle = \langle \Delta N_{i}^{*}  \Delta N_{j}^{*} \rangle + \sum_{R} \left( \langle N_{R} \rangle \langle \Delta n_i \Delta n_j \rangle_{R} + \langle (\Delta N_{R})^2 \rangle  \langle n_i \rangle_{R}  \langle n_j \rangle_{R} \right) .
\end{eqnarray} 

The third and fourth moments of multiplicity distribution can be obtained in a similar way, which read
\begin{eqnarray}
\langle \Delta N_i  \Delta N_j \Delta N_k \rangle
=&&\langle \Delta N_{i}^{*}  \Delta N_{j}^{*} \Delta N_{k}^{*} \rangle + \sum_{R} \langle N_R \rangle \langle \Delta n_i \Delta n_j \Delta n_k \rangle_R \nonumber\\
+&& \langle (\Delta N_{R})^{2} \rangle \left( \langle \Delta n_i \Delta n_j \rangle_{R} \langle n_k \rangle_{R}+ \langle \Delta n_i \Delta n_k \rangle_{R} \langle n_j \rangle_{R} + \langle \Delta n_k \Delta n_j \rangle_{R} \langle n_i \rangle_{R} \right) \nonumber\\
+&& \langle (\Delta N_{R})^3 \rangle \langle n_i \rangle_{R} \langle n_j \rangle_{R} \langle n_k \rangle_{R},
\end{eqnarray}

\begin{eqnarray}
C_4(N_i,N_j,N_k,N_l)=&& C_4(N_{i}^{*},N_{j}^{*},N_{k}^{*},N_{l}^{*})+ \sum_{R} \langle N_R \rangle \langle \Delta n_i \Delta n_j \Delta n_k \Delta n_l \rangle_R
\nonumber\\
-&& \langle N_{R} \rangle \left(\langle \Delta n_i \Delta n_j \rangle_{R} \langle \Delta n_k \Delta n_l \rangle_{R}+ \langle \Delta n_i \Delta n_k  \rangle_{R} \langle \Delta n_j \Delta n_l \rangle_{R}\right)
\nonumber\\
-&& \langle N_{R} \rangle \left(\langle \Delta n_i \Delta n_l \rangle_{R} \langle \Delta n_k \Delta n_j \rangle_{R} \right)
\nonumber\\
+&& \langle (\Delta N_{R})^{2} \rangle \left(\langle \Delta n_i \Delta n_j \Delta n_k \rangle_{R} \langle n_l \rangle_{R}+ \langle \Delta n_i \Delta n_j \Delta n_l \rangle_{R} \langle n_k \rangle_{R}\right)
\nonumber\\
+ &&\langle (\Delta N_{R})^{2} \rangle \left( \langle \Delta n_i \Delta n_k \Delta n_l \rangle_{R} \langle n_j \rangle_{R} + \langle \Delta n_k \Delta n_j \Delta n_l \rangle_{R} \langle n_i \rangle_{R} \right)
\nonumber\\
+&& \langle (\Delta N_{R})^{2} \rangle \left(\langle \Delta n_i \Delta n_j \rangle_{R} \langle \Delta n_k \Delta n_l \rangle_{R}+ \langle \Delta n_i \Delta n_k  \rangle_{R} \langle \Delta n_j \Delta n_l \rangle_{R}\right)
\nonumber\\
+&& \langle (\Delta N_{R})^{2} \rangle \left(\langle \Delta n_i \Delta n_l \rangle_{R} \langle \Delta n_k \Delta n_j \rangle_{R} \right)
\nonumber\\
+&& \langle (\Delta N_{R})^{3} \rangle \left(\langle \Delta n_i \Delta n_j \rangle_{R} \langle  n_k \rangle_{R} \langle  n_l \rangle_{R}+ \langle \Delta n_i \Delta n_k  \rangle_{R} \langle  n_j \rangle_{R} \langle  n_l \rangle_{R}\right)
\nonumber\\
+&& \langle (\Delta N_{R})^{3} \rangle \left(\langle \Delta n_i \Delta n_l \rangle_{R} \langle  n_j \rangle_{R} \langle  n_k \rangle_{R}+ \langle \Delta n_j \Delta n_k  \rangle_{R} \langle  n_i \rangle_{R} \langle  n_l \rangle_{R}\right)
\nonumber\\
+&& \langle (\Delta N_{R})^{3} \rangle \left(\langle \Delta n_j \Delta n_l \rangle_{R} \langle  n_i \rangle_{R} \langle  n_k \rangle_{R}+ \langle \Delta n_k \Delta n_l  \rangle_{R} \langle  n_i \rangle_{R} \langle  n_j \rangle_{R}\right)
\nonumber\\
+&& \left( \langle (\Delta N_{R})^4 \rangle - 3 \langle (\Delta N_{R})^2 \rangle^2 \right)\langle n_i \rangle_{R} \langle n_j \rangle_{R} \langle n_k \rangle_{R} \langle n_l \rangle_{R},
\end{eqnarray}
where the term $C_4$ on both sides of the equality is defined to be
\begin{eqnarray}
C_4(X_i,X_j,X_k,X_l)= && \langle \Delta X_i  \Delta X_j \Delta X_k \Delta X_l \rangle -\langle \Delta X_i  \Delta X_j \rangle \langle \Delta X_k  \Delta X_l \rangle \nonumber\\
- && \langle \Delta X_i  \Delta X_l \rangle \langle \Delta X_j  \Delta X_l \rangle - \langle \Delta X_i  \Delta X_k \rangle \langle \Delta X_j  \Delta X_l \rangle.
\end{eqnarray}

\bibliographystyle{h-physrev}
\bibliography{references_qian,references_ma,references_others}

\end{document}